**A novel experimental method of estimating tracer and intrinsic diffusion coefficients from multicomponent diffusion profiles**

Neelamegan Esakkiraja[1], Anuj Dash[1], Avik Mondal[1,2], K.C. Hari Kumar[3] and Aloke Paul[1*]
[1]Department of Materials Engineering, Indian Institute of Science, Bengaluru
[2]Research & Development Division, Tata Steel Ltd., Jamshedpur, Jharkhand, India
[3]Metallurgical and Materials Engineering, Indian Institute of Technology Madras, Chennai, India
Corresponding author: aloke@iisc.com, aloke.paul@gmail.com

## Abstract

A few decades earlier, Kirkaldy and Lane proposed an indirect method of estimating the tracer and intrinsic diffusion coefficients in a ternary system (without showing experimental verification), which is otherwise impossible following the Kirkendall marker experiments. Subsequently, Manning proposed the relations between the tracer and intrinsic diffusion coefficients in the multicomponent system by extending the Onsager formalism (although could not be estimated by intersecting the diffusion couples). By solving these issues in this article, we have now proposed the equations and method for estimating these parameters in pseudo-ternary diffusion couples in which diffusion paths can be intersected in multicomponent space. We have chosen NiCoFeCr system for verification of this method because of the availability of good quality diffusion couple experiments and estimated tracer diffusion coefficients of all the components measured by the radiotracer method. An excellent match is found when the tracer diffusion coefficients estimated following the newly proposed method are compared with the data estimated following the radiotracer method. Following, the intrinsic diffusion coefficients are estimated experimentally in a multicomponent system for the first time highlighting diffusional interactions between the components. We have further shown that the intrinsic diffusion coefficients are the same (if the vacancy wind effect is negligible/neglected) when estimated from other types of diffusion couples (pseudo-binary and body diagonal) in the same multi-component system. This method can be now extended to the Al, Ga, Si containing systems in which the estimation of tracer diffusion coefficients following the radiotracer method is difficult/impossible because of various reasons.

Keywords: Multicomponent diffusion; Interdiffusion; High Entropy Alloys; NiCoFeCr

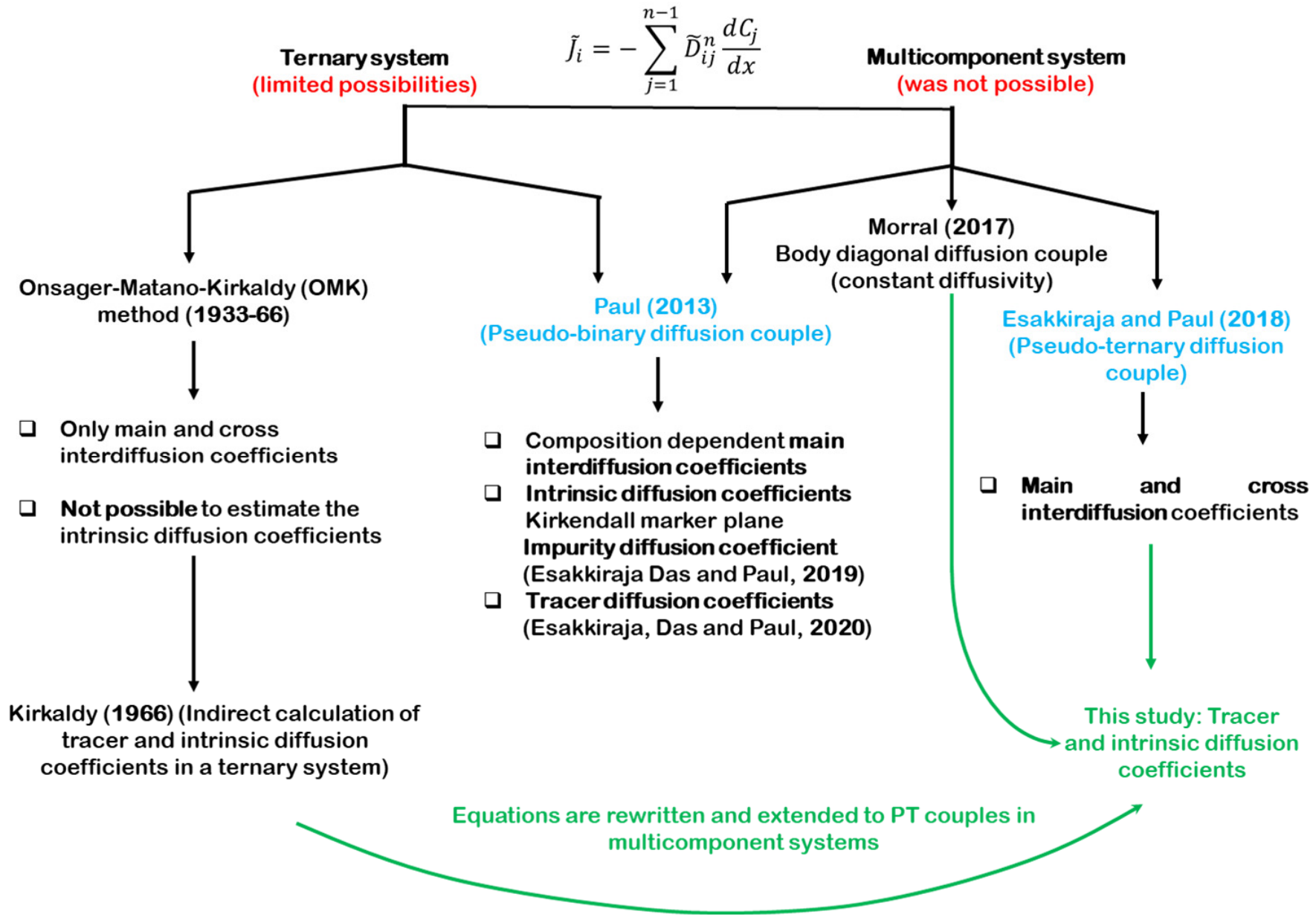

## 1. Introduction

The diffusion community suffered during the last many decades because of the lack of experimental methods for the estimation of the diffusion coefficients in inhomogeneous multicomponent systems. All types of diffusion coefficients could be estimated only in binary systems, whereas, only the interdiffusion coefficients could be estimated in ternary systems. No diffusion parameters could be estimated from interdiffusion profiles in a system with more than three components until recently. This led to an unbridgeable gap between the fundamental studies conducted in simpler systems and the need for understanding the diffusion phenomenon in various multicomponent systems in applications.

The newly proposed pseudo-binary (PB) and pseudo-ternary (PT) diffusion couple methods in the multicomponent system by Paul and his co-workers [1-4] have solved this problem of the last many decades. One can intersect the diffusion paths by producing the PT diffusion couples, which is otherwise difficult (or maybe impossible) [4]. One may compromise the restriction by estimating the diffusion coefficients following the concept of body diagonal method [5] in which the diffusion paths may not intersect but pass closely [6]. At present, the PB and PT methods are used by different groups in various systems because of the relative ease of designing experiments and subsequent analysis [7-13]. Sometimes the pseudo-binary diffusion couples were produced but the diffusion coefficients were estimated differently (following simulation/numerical methods) instead of a direct calculation from the composition profiles [14, 15]. With the help of the reliable thermodynamic database, one can identify the composition range in which one can successfully produce the ideal or near-ideal PB and PT diffusion couples. However, it is not always easy to find reliable databases to do calculations in multicomponent systems in the composition range of interest. Sometimes, thermodynamic information obtained from databank systems such as Thermo-Calc pertaining to systems comprising multi-principal alloys should be treated with caution [16]. A major mismatch was found between experimentally produced diffusion profiles and simulated profiles utilizing thermodynamic databases in combination with the different kinetic databases in NiCoFeCrMn system [17]. In such a situation, the possibilities of finding PB and PT couples can be found experimentally based on simple trial experiments, which will be discussed in detail in future. The experimentally estimated diffusion coefficients following these methods can be now used to verify the data generated by newly proposed numerical methods [18]. The ideal PB and PT profiles may not be possible to produce in the whole composition range of a multicomponent system. By correlating the experimentally estimated diffusion coefficients with numerical methods in a composition range where PB and PT couples can be produced, one can extend these analyses to other composition ranges where we cannot produce such profiles. Recently, the radiotracer method was combined with the diffusion couple method for estimation of the composition-dependent diffusion coefficients instead of estimating at one composition from a single experiment [19-21]. This may become even more versatile if it is combined with PB and PT diffusion couples in a multicomponent system in which ideal/near-ideal diffusion profiles are produced. This is done earlier but in a PB couple with major non-ideality, which should be treated as a conventional diffusion couple in which all the components produce the diffusion profiles [17].

Irrespective of the recent developments, a very important aspect could not be fulfilled until now based on a purely experimental analysis in a multicomponent system. The intrinsic diffusion coefficients, which are fundamental diffusion coefficients of components under the influence of thermodynamic driving forces, can be estimated following the Kirkendall marker experiment only in binary diffusion couples [22, 23] and now also in PB multicomponent diffusion couples [4]. However, only the main intrinsic diffusion coefficients can be estimated in a PB diffusion couple, which is not enough to understand the complicated diffusion process without knowing the cross intrinsic diffusion coefficients. Dayananda [24], Dayananda and Whittle [25] attempted to estimate the mobilities differently and demonstrated these in ternary systems. The main as well as the cross intrinsic diffusion coefficients cannot be estimated in ternary or multicomponent PT diffusion couples following the Kirkendall marker experiment. It needs the presence of the Kirkendall marker plane at the composition of intersection in all the diffusion couples, which cannot be achieved experimentally unless found incidentally without prior knowledge of the diffusion matrix. Therefore, the diffusion process is explained based on the estimated interdiffusion coefficients, which are kind of average of certain sets of intrinsic diffusion coefficients in a multicomponent system. The possibility of estimation of the intrinsic diffusion coefficients experimentally can enhance the understanding of the diffusion process dramatically in a multicomponent system. Additionally, there is a need for creating a link between the tracer and intrinsic diffusion coefficients to the interdiffusion coefficients in the PT couples to understand the phenomenological diffusion process in this type of diffusion couple. In this article, we explore the possibilities of estimating both tracer and intrinsic diffusion coefficients from the PT diffusion couples in the multicomponent system.

As already mentioned, we can now estimate the interdiffusion coefficients by intersecting the diffusion couples in multicomponent space utilizing the concept of PT diffusion couples [2-4]. In this study, we first propose a novel approach of estimating the tracer diffusion coefficients from the estimated interdiffusion coefficients at the intersection of two PT diffusion couples by extending the method proposed by Kirkaldy and Lane [26] and van Loo et al. [27, 28] in a ternary system after rewriting several equations fulfilling the condition of PT diffusion profiles in a multicomponent system. To the best of our knowledge, the relations proposed by Kirkaldy and Lane in the ternary system was never followed because of the unavailability of reliable thermodynamic data [26, 27] during that period. The group of van Loo estimated only the tracer diffusion coefficients from the interdiffusion coefficients in a ternary system [27]. In the process of estimating the intrinsic and tracer diffusion coefficients, we establish a link between the relations proposed in conventional diffusion couples by Kirkaldy et al. [26, 29, 30] and Manning [31, 32] based on the Onsager formalism [33-35] to the PT diffusion couples in a multicomponent system. Equations are derived for both the situations i.e. by neglecting and considering the vacancy wind effects.

The data estimated following this PT diffusion couple method using the newly proposed relations are compared with the tracer diffusion coefficients measured by the radiotracer technique to find a very good match. Therefore, now, one can first estimate the tracer diffusion coefficients of all the components following the radio-tracer method or the pseudo-binary and pseudo-ternary diffusion couple methods. Following, utilizing the relations proposed, one can calculate all types of intrinsic and interdiffusion coefficients in a multicomponent system. Therefore, there is no need to conduct experiments following the conventional method since anyway we cannot estimate these coefficients by intersecting the diffusion couples in a multi-component space. At present, only a very few groups have the facility for measuring tracer diffusion coefficients using radioisotopes because of stringent restrictions of maintaining the facilities. On the other hand, the diffusion couple methods are easier to practice. Therefore, the analysis described in this study will open various new possibilities.

## 2. Relations for estimation of tracer, intrinsic and interdiffusion coefficients from PT diffusion couple

The relations for estimating the interdiffusion coefficients in a PT couple are described earlier [3-4] (see supplementary file). For the sake of continuity, we introduce these relations before proposing the relations in correlation to the tracer and intrinsic diffusion coefficients.

The interdiffusion flux ($\tilde{J}_i$) and the interdiffusion coefficients ($\widetilde{D}_{ij}^n$) in a *n* component system are related considering a constant molar volume by [22-23]

$$\tilde{J}_i = -\sum_{j=1}^{n-1} \widetilde{D}_{ij}^n \frac{\partial C_j}{\partial x} = -\frac{1}{V_m} \sum_{j=1}^{n-1} \widetilde{D}_{ij}^n \frac{\partial N_j}{\partial x} \quad (1a)$$

where $\widetilde{D}_{ij}^n = \widetilde{D}_{ij} - \widetilde{D}_{in}$ (see supplementary file). Component *n* is the dependent variable. $\widetilde{D}_{ii}^n$ is the main interdiffusion coefficient for component *i*, which is related to its composition gradient and $\widetilde{D}_{ij}^n$ is the cross interdiffusion coefficient, which is related to the composition gradient of another component *j*. These cross-terms are important for understanding the complex diffusional interactions between the components [6]. *x* is the position parameter. Further, the compositions of different components are related by $\sum_{i=1}^{n} N_i = 1$. Therefore, the interdiffusion fluxes are related by

$$\sum_{i=1}^{n} \tilde{J}_i = 0 \tag{1b}$$

This leads to the estimation of the interdiffusion coefficients with one component (let us say component *n* in the above equation) as the dependent variable since we have (*n-1*) independent interdiffusion fluxes (see supplementary file). The interdiffusion flux of component *i* can be calculated from the composition profile using [22, 23]

$$V_m \tilde{J}_i = -\frac{N_i^+ - N_i^-}{2t}\left[(1 - Y_i^*)\int_{x^{-\infty}}^{x^*} Y_i dx + Y_i^* \int_{x^*}^{x^{+\infty}} (1 - Y_i) dx\right] \tag{1c}$$

where $Y_i = \frac{N_i^* - N_i^-}{N_i^+ - N_i^-}$ is the Sauer-Freise composition normalization variable [36]. $N_i^-$ and $N_i^+$ are the compositions of unaffected left and right-hand ends of the diffusion couple.

In lattice fixed frame of reference, the intrinsic fluxes are related to the intrinsic diffusion coefficients by

$$J_i = -\sum_{j=1}^{n} D_{ij}\frac{\partial C_j}{\partial x} = -\sum_{j=1}^{n-1} D_{ij}^n \frac{\partial C_j}{\partial x} = -\frac{1}{V_m}\sum_{j=1}^{n-1} D_{ij}^n \frac{\partial N_j}{\partial x} \tag{1d}$$

The intrinsic fluxes of different components are related to the vacancy flux $(J_v)$ by

$$J_1 + J_2 + J_3 + \dots J_n + J_v = 0 \tag{1e}$$

and the interdiffusion and intrinsic fluxes are related by

$$\tilde{J}_i = J_i - N_i \sum_{k=1}^{n} J_k \tag{1f}$$

Using Equation 1f and comparing Equations 1a and 1d, the interdiffusion and intrinsic diffusion coefficients are related by

$$\tilde{D}_{ij}^n = D_{ij}^n - N_i\left(\sum_{k=1}^{n} D_{kj}^n\right) \tag{1g}$$

The intrinsic flux of a component at the Kirkendall marker composition can be calculated from (see supplementary file)

$$V_m J_i = -\frac{1}{2t}\left[N_i^+ \int_{x^{-\infty}}^{x_K} Y_i dx - N_i^- \int_{x_K}^{x^{+\infty}} (1 - Y_i) dx\right] \tag{1h}$$

If a component, which is kept with the same composition in both the ends of a diffusion couple, remain constant in the whole interdiffusion zone, the interdiffusion and intrinsic diffusion flux will be zero. This can be understood from Equation 1c and 1h. If a component indeed remains constant without developing the diffusion profiles in an interdiffusion zone, both $\int_{x^{-\infty}}^{x^*} Y_i dx$ and $\int_{x^*}^{x^{+\infty}} (1 - Y_i) dx$ will be zero. In a non-ideal condition, one component which is kept with the same composition in two ends of the diffusion couple may develop the diffusion profile and the interdiffusion flux of that component will not be zero. In such a situation, the interdiffusion flux cannot be calculated utilizing Equation 1c because of the term $(N_i^+ - N_i^-)$, which will otherwise calculate the interdiffusion flux as zero wrongly. Therefore, we should use the relation proposed by Matano-Kirkaldy, as given in the supplementary file. In this case, the diffusion couple should not be treated or even named as PB or PT diffusion couples, since in a PB couple only two components and a PT couple only three components should develop the diffusion profiles keeping all other components constant. One may still consider a very small non-ideality for calculation without introducing mentionable error i.e. when the composition profile is found to deviate within the error of composition measurement of Electron Probe Micro Analyzer from the constant average value.

Since only three components (let us say components 1, 2 and 3) develop the diffusion profiles in an ideal PT diffusion couple keeping all other components (4 to n) constant and, further, we have also $\tilde{J}_1 + \tilde{J}_2 + \tilde{J}_3 = 0$, because $\tilde{J}_4 = \cdots .. \tilde{J}_n = 0$. Therefore, Equations 1a and 1b expressed with compositions $N_i$ should be expressed with modified compositions $M_i^T$ fulfilling this condition such that

$$N_1 + N_2 + N_3 = 1 - N_4 \ldots - N_n = 1 - N_f = N_v \tag{2a}$$

$$\frac{N_1}{N_v} + \frac{N_2}{N_v} + \frac{N_3}{N_v} = 1 \tag{2b}$$

$$M_1^T + M_2^T + M_3^T = 1 \tag{2c}$$

where $N_f = N_4 \ldots + N_n$ is the sum of the composition of components which remain constant (fixed) and $N_v = N_1 + N_2 + N_3$ is the sum of the composition of components which contribute to the development of interdiffusion profiles. It should be noted here that the components which are kept constant do not contribute to the interdiffusion flux but take part in the redistribution of the components. Therefore, the composition profiles ($N_i$ vs. *x*) should be converted first to modified composition profiles ($M_i^T$ vs. *x*). The modified compositions should be expressed differently in an intermetallic compound compared to solid solutions, which is discussed in Ref. [7] (see supplementary file). Equation 1a in PT couple considering component 1 as a dependent can be expressed with modified composition profiles as (see supplementary file)

$$\tilde{J}_i = -\frac{1}{V_m}\sum_{j=2}^{3} \tilde{D}_{ij}^{1} \frac{\partial M_j^T}{\partial x} \tag{3a}$$

This can be expanded to

$$\tilde{J}_1 = -(\tilde{J}_2 + \tilde{J}_3) \tag{3b}$$

$$V_m \tilde{J}_2 = -\tilde{D}_{22}^{1} \frac{\partial M_2^T}{\partial x} - \tilde{D}_{23}^{1} \frac{\partial M_3^T}{dx} \tag{3c}$$

$$V_m \tilde{J}_3 = -\tilde{D}_{32}^{1} \frac{\partial M_2^T}{dx} - \tilde{D}_{33}^{1} \frac{\partial M_3^T}{dx} \tag{3d}$$

The interdiffusion flux can be expressed as (see supplementary file)

$$V_m \tilde{J}_i = -\frac{M_i^{T+} - M_i^{T-}}{2t}\left[\left(1 - Y_{i,T}^*\right)\int_{x^{-\infty}}^{x^*} Y_{i,T}\,dx + Y_{i,T}^*\int_{x^*}^{x^{+\infty}} (1 - Y_{i,T})\,dx\right] \tag{3e}$$

where $Y_{i,T} = \frac{M_i^{T*} - M_i^{T-}}{M_i^{T+} - M_i^{T-}}$ is the modified composition normalization variable.

Wrong value of the interdiffusion flux will be calculated if Equation 1c is used instead of 3e since the sum of compositions of the components which produce the diffusion profiles at any location i.e. $N_1 + N_2 + N_3 \neq 1$ leading to wrongly $\tilde{J}_1 + \tilde{J}_2 + \tilde{J}_3 \neq 0$.

The interdiffusion and intrinsic fluxes in a PT couple are related by (see supplementary file)

$$\tilde{J}_i = J_i - M_i^T \sum_{k=1}^{n} J_k \tag{4}$$

In lattice fixed frame of reference, the intrinsic flux of component *i* ($J_i$) is related to the intrinsic diffusion coefficients ($D_{ij}^1$) with respect to the modified composition profile. Considering component 1 as a dependent, this can be expressed as

$$J_i = -\frac{1}{V_m}\sum_{j=2}^{3} D_{ij}^{1} \frac{\partial M_j^T}{\partial x} \tag{5a}$$

Equation 5a can be expanded to

$$V_m J_1 = -D_{12}^1 \frac{\partial M_2^T}{dx} - D_{13}^1 \frac{\partial M_3^T}{dx} \quad (5b)$$

$$V_m J_2 = -D_{22}^1 \frac{\partial M_2^T}{dx} - D_{23}^1 \frac{\partial M_3^T}{dx} \quad (5c)$$

$$V_m J_3 = -D_{32}^1 \frac{\partial M_2^T}{dx} - D_{33}^1 \frac{\partial M_3^T}{dx} \quad (5d)$$

The intrinsic fluxes of different components are related to the vacancy flux by

$$J_1 + J_2 + J_3 + J_v = 0; \qquad J_4 = \cdots = J_n = 0 \quad (5e)$$

where the Kirkendall marker velocity $v_K$ is related to the vacancy flux $J_v$ and the molar volume $V_M$.

The Darken equation [37] in a PT couple can be expressed with respect to the modified composition as

$$\tilde{J}_i = J_i + v_K \frac{M_i^T}{V_m} \quad (5f)$$

Taking as the sum of the above equation for all the components (i.e. 1, 2 and 3, which produce the diffusion profiles) and from Equations 1b and 2c, we have

$$v_K = -V_m \left(J_1 + J_2 + J_3\right) = V_m J_v \quad (5g)$$

Comparing Equations 3, 4 and 5, the interdiffusion and intrinsic diffusion coefficients in a PT couple can be related as (see supplementary file)

$$\widetilde{D}_{ij}^n = D_{ij}^n - M_i^T \left(\sum_{k=1}^n D_{kj}^n\right) \ (i = 1, 2, 3) \quad (6a)$$

This can be expanded to relate the four interdiffusion coefficients with six intrinsic diffusion coefficients in a PT couple as

$$\widetilde{D}_{22}^1 = D_{22}^1 - M_2^T (D_{12}^1 + D_{22}^1 + D_{32}^1) \quad (6b)$$

$$\widetilde{D}_{23}^1 = D_{23}^1 - M_2^T (D_{13}^1 + D_{23}^1 + D_{33}^1) \quad (6c)$$

$$\widetilde{D}_{32}^1 = D_{32}^1 - M_3^T (D_{12}^1 + D_{22}^1 + D_{32}^1) \quad (6d)$$

$$\widetilde{D}_{33}^1 = D_{33}^1 - M_3^T (D_{13}^1 + D_{23}^1 + D_{33}^1) \quad (6e)$$

The interdiffusion coefficients can be estimated at the composition of the intersection of two PT diffusion couples in a multi-component space [2-4] by writing the four equations (following Equation 3) for the estimation of four independent interdiffusion coefficients. However, it is almost impossible to estimate the intrinsic diffusion coefficients since the mandatory condition of finding the Kirkendall marker plane at the same composition of intersection is almost impossible to achieve without prior knowledge of diffusion matrix over the composition range of the diffusion couples and then predict the end-member composition accordingly. Unless the marker plane is found at the composition of the intersection, six equations from two diffusion couples at the same composition cannot be written for the estimation of the six independent intrinsic diffusion coefficients. This is the reason that these parameters could not be estimated even in a ternary system unless found accidentally.

In the meantime, Kirkaldy and Lane [26] advocated the use of an indirect method in a ternary system following which one can estimate first the tracer diffusion coefficients at the composition of the intersection of two diffusion couples utilizing the thermodynamic parameters. Following that procedure, one can estimate the intrinsic diffusion coefficients. However, they never used this method for the estimation of these data because of the lack of availability of reliable thermodynamic data in a ternary system during that era [26]. Much later, the group of van Loo followed a similar approach for estimation of the tracer diffusion coefficients [27]. However, they did not calculate the intrinsic diffusion coefficients.

In this article, we adopt this concept proposed in a ternary system by extending it to PT diffusion couples in a multicomponent system. Therefore, we first need to establish the relations in the PT diffusion couple in which only three components develop the diffusion profiles keeping other components constant. Moreover, instead of calculating these parameters with respect to the chemical potential gradients, which was proposed by Kirkaldy and Lane [26], we express the relations with respect to the thermodynamic factors as proposed by van Loo et al. [28] since these are material constants and help in reducing uncertainty/error in calculation. Both Kirkaldy and Lane and the group of van Loo developed the relations neglecting the contribution of the vacancy wind effect as proposed by Manning [32]. We aim at establishing the relations for both neglecting and considering the vacancy wind effect in a PT couple by modifying the equations proposed by Manning in a multicomponent system.

The intrinsic flux can be related to the chemical potential gradient $\frac{\partial \mu_j}{\partial x}$ and the phenomenological constants $L_{ij}$ by [29, 33-35]

$$V_m J_i = -\sum_{j=1}^{n} L_{ij} \frac{\partial \mu_j}{\partial x} \tag{7}$$

Manning [32] proposed the relations between the intrinsic and tracer diffusion coefficients considering vacancy wind effect in multicomponent diffusion profile, which can be expressed for component *i* considering constant molar volume as

$$V_m J_i = -\frac{N_i D_i^*}{RT} \frac{\partial \mu_i}{dx} - \beta N_i D_i^* V_m J_v \tag{8a}$$

where $\beta = \frac{2}{(S_o+2)\sum_{j=1}^{n} N_j D_j^*}$ and $S_o$ is the structure factor. In FCC crystal the value of $S_o$ is 7.15 [31].

This applies to the conventional diffusion couple in which all the components develop the diffusion profiles. Taking a sum over all the components and since $J_v = -\sum_{j=1}^{n} J_j$ [22, 23, 26], Equation 8a can be expressed as [32]

$$V_m J_i = -\frac{N_i D_i^*}{RT} \frac{\partial \mu_i}{dx} - \sum_{j=1}^{n} \frac{N_j D_j^*}{RT} (\xi N_i D_i^*) \frac{\partial \mu_j}{dx} \tag{8b}$$

where $\xi = \frac{\beta}{1-\beta\sum_{j=1}^{n} N_j D_j^*} = \frac{2}{S_o \sum_{j=1}^{n} N_j D_j^*}$

Le Claire [38] and Kirkaldy and Lane [26] considered only the first part of Equations 8a and 8b. Other terms with $\xi$ come from the consideration of the cross phenomenological constants contributing to the vacancy wind effect. In a PT diffusion couple, in which only three components develop the diffusion profiles keeping other components as the constants, Equation 8a with respect to the modified compositions can be expressed as

$$V_m J_i = -\frac{M_i^T D_i^*}{RT} \frac{\partial \mu_i}{dx} - \beta_{PT} M_i^T D_i^* V_m J_v \tag{9a}$$

where $\beta_{PT} = \frac{2}{(S_o+2)\sum_{j=1}^{3} M_j^T D_j^*}$

First Equation 9a is expressed for intrinsic fluxes of all the components *i.e.* $J_1$, $J_2$ and $J_3$. After taking sum and then by replacing $(J_1 + J_2 + J_3) = -J_v$ (see Equation 5e), we can derive the relation of $J_v$. Replacing this in Equation 9a, we have

$$V_m J_i = -\frac{M_i^T D_i^*}{RT}\frac{\partial \mu_i}{dx} - \sum_{j=1}^{3}\frac{M_j^T D_j^*}{RT}\left(\xi_{PT} M_i^T D_i^*\right)\frac{\partial \mu_j}{dx} \quad (9b)$$

where $\xi_{PT} = \frac{\beta_{PT}}{1-\beta_{PT}\sum_{j=1}^{3} M_j^T D_j^*} = \frac{2}{S_o \sum_{j=1}^{3} M_j^T D_j^*}$

**2.1 The relations neglecting the contribution of cross phenomenological constants and vacancy wind effect in PT diffusion couples**

Kirkaldy and Lane [26] in support to Le Claire [38] argued that the off-diagonal phenomenological terms in Equation 7 could be approximated as zero in substitutional diffusion controlled by vacancy mechanism. On the other hand, Manning considered these cross terms by introducing the concept of the vacancy wind effect [32]. In several binary systems, indeed the contribution of this vacancy wind effect is found to be negligible [22]. This is the reason that the existence of the vacancy wind effect could not be verified experimentally by comparing the interdiffusion/intrinsic diffusion coefficients and the tracer diffusion coefficients although this effect is indeed present [22]. The contribution to the PT diffusion couples should be discussed, which may play a significant role in certain systems. Let us first discuss the relations neglecting the vacancy wind effect since this is followed in various numerical and simulation methods. Following this, we express the relations considering the vacancy wind effect.

Equation 9b by neglecting the vacancy wind effect (i.e. by neglecting all the terms containing $\xi$) can be expressed as

$$V_m J_i = -\frac{M_i^T D_i^*}{RT}\frac{\partial \mu_i}{\partial x} \quad (10a)$$

Following Onsager [35] since only (*n-1*) compositions vary independently in a *n* component system [29] and treat composition of component 1 as dependent, we can express

$$\frac{\partial \mu_i}{\partial x} = \frac{\partial \mu_i}{\partial N_2}\frac{\partial N_2}{\partial x} + \frac{\partial \mu_i}{\partial N_3}\frac{\partial N_3}{\partial x} + \frac{\partial \mu_i}{\partial N_4}\frac{\partial N_4}{\partial x} + \cdots + \frac{\partial \mu_i}{\partial N_n}\frac{\partial N_n}{\partial x} \quad (10b)$$

In a PT couple, since only components 1, 2 and 3 develop the diffusion profiles $\left(\frac{\partial N_4}{\partial x} = \cdots \frac{\partial N_n}{\partial x} = 0\right)$, we have

$$\frac{\partial \mu_i}{\partial x} = \frac{\partial \mu_i}{\partial N_2}\frac{\partial N_2}{\partial x} + \frac{\partial \mu_i}{\partial N_3}\frac{\partial N_3}{\partial x} \quad (10c)$$

We should realize that the chemical potential or the activity gradient of a component in different diffusion couples at the composition of the intersection will be different depending on the developed composition profiles. Therefore, for the sake of ease of analysis and to correlate the diffusion coefficients in different types of diffusion couples, we express these relations using the concept of thermodynamic factor. Equation 10c can be rewritten with respect to activity ($a_i$) of component *i* (since $\mu_i = \mu_i^o + RT ln a_i$) [28] as

$$\frac{\partial \mu_i}{\partial x} = \frac{1}{N_2}\frac{RT\partial ln a_i}{\partial ln N_2}\frac{\partial N_2}{\partial x} + \frac{1}{N_3}\frac{RT\partial ln a_i}{\partial ln N_3}\frac{\partial N_3}{\partial x}$$

$$\frac{\partial \mu_i}{\partial x} = \frac{RT}{N_2}\phi_{i2}^1\frac{\partial N_2}{\partial x} + \frac{RT}{N_3}\phi_{i3}^1\frac{\partial N_3}{\partial x} \quad (10d)$$

where $\emptyset_{ij}^1 = \left(\frac{\partial ln a_i}{\partial ln N_j}\right)_{p,T,\Sigma}$ is the thermodynamic factor. The subscript $i$ and $j$ in $\emptyset_{ij}^1$ can vary from 2 to *n* and $\Sigma$ indicates that the derivative is taken after eliminating $N_1$ in the expression for $ln a_i$.

Dividing by $N_v$ $(= N_1 + N_2 + N_3)$ in both numerator and denominator in the right-hand side of the equation 10d, we have

$$\frac{\partial \mu_i}{\partial x} = \frac{RT}{M_2^T}\phi_{i2}^1 \frac{\partial M_2^T}{\partial x} + \frac{RT}{M_3^T}\phi_{i3}^1 \frac{\partial M_3^T}{\partial x} \tag{10e}$$

Therefore, Equation 10a can be rewritten with respect to the thermodynamic factors for different components developing the diffusion profiles and can be expressed as

$$V_m J_1 = -\frac{M_1^T D_1^*}{RT}\frac{\partial \mu_1}{\partial x} = -\frac{M_1^T D_1^*}{RT}\left(\frac{RT}{M_2^T}\phi_{12}^1 \frac{\partial M_2^T}{\partial x} + \frac{RT}{M_3^T}\phi_{13}^1 \frac{\partial M_3^T}{\partial x}\right) \tag{11a}$$

$$V_m J_2 = -\frac{M_2^T D_2^*}{RT}\frac{\partial \mu_2}{\partial x} = -\frac{M_2^T D_2^*}{RT}\left(\frac{RT}{M_2^T}\phi_{22}^1 \frac{\partial M_2^T}{\partial x} + \frac{RT}{M_3^T}\phi_{23}^1 \frac{\partial M_3^T}{\partial x}\right) \tag{11b}$$

$$V_m J_3 = -\frac{M_3^T D_3^*}{RT}\frac{\partial \mu_3}{\partial x} = -\frac{M_3^T D_3^*}{RT}\left(\frac{RT}{M_2^T}\phi_{32}^1 \frac{\partial M_2^T}{\partial x} + \frac{RT}{M_3^T}\phi_{33}^1 \frac{\partial M_3^T}{\partial x}\right) \tag{11c}$$

Comparing Equations 5 and 11, we have

$$-D_{12}^1 \frac{\partial M_2^T}{\partial x} - D_{13}^1 \frac{\partial M_3^T}{\partial x} = -\frac{M_1^T D_1^*}{RT}\left(\frac{RT}{M_2^T}\phi_{12}^1 \frac{\partial M_2^T}{\partial x} + \frac{RT}{M_3^T}\phi_{13}^1 \frac{\partial M_3^T}{\partial x}\right) \tag{12a}$$

$$-D_{22}^1 \frac{\partial M_2^T}{\partial x} - D_{23}^1 \frac{\partial M_3^T}{\partial x} = -\frac{M_2^T D_2^*}{RT}\left(\frac{RT}{M_2^T}\phi_{22}^1 \frac{\partial M_2^T}{\partial x} + \frac{RT}{M_3^T}\phi_{23}^1 \frac{\partial M_3^T}{\partial x}\right) \tag{12b}$$

$$-D_{32}^1 \frac{\partial M_2^T}{\partial x} - D_{33}^1 \frac{\partial M_3^T}{\partial x} = -\frac{M_3^T D_3^*}{RT}\left(\frac{RT}{M_2^T}\phi_{32}^1 \frac{\partial M_2^T}{\partial x} + \frac{RT}{M_3^T}\phi_{33}^1 \frac{\partial M_3^T}{\partial x}\right) \tag{12c}$$

Comparing the coefficients of the differentials, the intrinsic and tracer diffusion coefficients in PT couple are related to the thermodynamic parameters by

$$D_{ii}^1 = D_i^* \phi_{ii}^1 \tag{13a}$$

$$D_{ij}^1 = \frac{M_i^T}{M_j^T} D_i^* \phi_{ij}^1 = \frac{N_i}{N_j} D_i^* \phi_{ij}^1 \tag{13b}$$

These are the same relations expressed (by neglecting the vacancy wind effect) when the data are estimated from conventional diffusion couples in a multicomponent system by producing the diffusion profiles of all components (see supplementary file for details) only if we could intersect the diffusion paths. Now let us relate the interdiffusion coefficients with respect to the tracer diffusion coefficients, which can be expressed by combining Equation 13 and Equation 6 as

$$\tilde{D}_{22}^1 = D_{22}^1 - M_2^T(D_{12}^1 + D_{22}^1 + D_{32}^1) = (1 - M_2^T)D_2^*\phi_{22}^1 - (M_1^T D_1^*\phi_{12}^1 + M_3^T D_3^*\phi_{32}^1) \tag{14a}$$

$$\tilde{D}_{23}^1 = D_{23}^1 - M_2^T(D_{13}^1 + D_{23}^1 + D_{33}^1) = \frac{M_2^T}{M_3^T}[(1 - M_2^T)D_2^*\phi_{23}^1 - (M_1^T D_1^*\phi_{13}^1 + M_3^T D_3^*\phi_{33}^1)] \tag{14b}$$

$$\tilde{D}_{32}^1 = D_{32}^1 - M_3^T(D_{12}^1 + D_{22}^1 + D_{32}^1) = \frac{M_3^T}{M_2^T}[(1 - M_3^T)D_3^*\phi_{32}^1 - (M_1^T D_1^*\phi_{12}^1 + M_2^T D_2^*\phi_{22}^1)] \tag{14c}$$

$$\tilde{D}_{33}^1 = D_{33}^1 - M_3^T(D_{13}^1 + D_{23}^1 + D_{33}^1) = (1 - M_3^T)D_3^*\phi_{33}^1 - (M_1^T D_1^*\phi_{13}^1 + M_2^T D_2^*\phi_{23}^1) \tag{14d}$$

Therefore, one can follow the steps for calculation of tracer and intrinsic diffusion coefficients at the intersection of two PT diffusion couples as:

(i) First, estimate the interdiffusion coefficients at the composition of the intersection of two diffusion couples directly from the calculated interdiffusion fluxes (see Equation 3).

(ii) Following, calculate the tracer diffusion coefficients utilizing the thermodynamic factors and by solving Equation 14. It should be noted here that since only three components develop the diffusion profiles, Equation 14 is related to tracer diffusion coefficients of three components. Therefore, we need three equations for these calculations at a time. One is suggested to take a different combination of three different equations in rotation and calculate the tracer diffusion coefficients. Following, take an average of the calculated tracer values and calculate back the interdiffusion coefficients to make sure that the back-calculated interdiffusion coefficients are similar to the values estimated directly from the diffusion profiles.

(iii) These equations are very sensitive to the experimental data and sometimes may produce an illogical value, which should not be considered for calculation of average tracer diffusion coefficients. This is not witnessed in our analysis in NiCoFeCr but may be witnessed in other systems depending on the quality of experiments, measurement of compositions and correctness of a calculation or even because of wrong thermodynamic parameters. The back-calculation of the interdiffusion coefficients from the average tracer diffusion coefficients as suggested in the previous point will instil confidence if the tracer diffusion coefficients are not already available to compare. In this study, we have considered the four-component NiCoFeCr system since we can compare the tracer diffusion coefficients calculated in this study with the data measured following the radiotracer method.

(iv) As a next step, we can calculate the intrinsic diffusion coefficients utilizing the thermodynamic parameters from Equation 13.

(v) One can even estimate the tracer and intrinsic diffusion coefficients directly from the calculated interdiffusion fluxes at the composition of intersection by replacing Equation 11 in Equation 4 instead of the calculation after estimating the interdiffusion coefficients. As discussed in the next section, it will help to extend this method by considering only two diffusion couples following the body diagonal method in a multicomponent system in which the diffusion couples may not intersect but may pass closely.

Now we discuss the relations considering the vacancy wind effect. Equation 9b in a PT diffusion couple considering component 1 as the dependent variable can be expressed as

$$V_m J_i = -\frac{M_i^T D_i^*}{RT}\frac{\partial \mu_i}{dx} - \frac{\xi_{PT} M_i^T D_i^*}{RT}\sum_{j=1}^{3} M_j^T D_j^* \frac{\partial \mu_j}{dx} \tag{15}$$

This is expressed for the three diffusing components as

$$V_m J_1 = -\frac{M_1^T D_1^*}{RT}\frac{\partial \mu_1}{dx} - \frac{\xi_{PT} M_1^T D_1^*}{RT}\left[M_1^T D_1^* \frac{\partial \mu_1}{dx} + M_2^T D_2^* \frac{\partial \mu_2}{dx} + M_3^T D_3^* \frac{\partial \mu_3}{dx}\right] \tag{16a}$$

$$V_m J_2 = -\frac{M_2^T D_2^*}{RT}\frac{\partial \mu_2}{dx} - \frac{\xi_{PT} M_2^T D_2^*}{RT}\left[M_1^T D_1^* \frac{\partial \mu_1}{dx} + M_2^T D_2^* \frac{\partial \mu_2}{dx} + M_3^T D_3^* \frac{\partial \mu_3}{dx}\right] \tag{16b}$$

$$V_m J_3 = -\frac{M_3^T D_3^*}{RT}\frac{\partial \mu_3}{dx} - \frac{\xi_{PT} M_3^T D_3^*}{RT}\left[M_1^T D_1^* \frac{\partial \mu_1}{dx} + M_2^T D_2^* \frac{\partial \mu_2}{dx} + M_3^T D_3^* \frac{\partial \mu_3}{dx}\right] \tag{16c}$$

Replacing, Equation 10e in Equation 15, we have

$$V_m J_i = -\frac{M_i^T D_i^*}{RT}\left(\frac{RT}{M_2^T}\phi_{i2}^1 \frac{\partial M_2^T}{\partial x} + \frac{RT}{M_3^T}\phi_{i3}^1 \frac{\partial M_3^T}{\partial x}\right) - \frac{\xi_{PT} M_i^T D_i^*}{RT}\left[\sum_{j=1}^{3} M_j^T D_j^* \left(\frac{RT}{M_2^T}\phi_{j2}^1 \frac{\partial M_2^T}{\partial x} + \frac{RT}{M_3^T}\phi_{j3}^1 \frac{\partial M_3^T}{\partial x}\right)\right] \tag{17}$$

These can be written for different components from Equation 16 as

$$V_m J_1 = -\frac{M_1^T D_1^*}{RT}\left(\frac{RT}{M_2^T}\phi_{12}^1 \frac{\partial M_2^T}{\partial x} + \frac{RT}{M_3^T}\phi_{13}^1 \frac{\partial M_3^T}{\partial x}\right) - \frac{\xi_{PT} M_1^T D_1^*}{RT}\left[M_1^T D_1^* \left(\frac{RT}{M_2^T}\phi_{12}^1 \frac{\partial M_2^T}{\partial x} + \frac{RT}{M_3^T}\phi_{13}^1 \frac{\partial M_3^T}{\partial x}\right) + M_2^T D_2^* \left(\frac{RT}{M_2^T}\phi_{22}^1 \frac{\partial M_2^T}{\partial x} + \frac{RT}{M_3^T}\phi_{23}^1 \frac{\partial M_3^T}{\partial x}\right) + M_3^T D_3^* \left(\frac{RT}{M_2^T}\phi_{32}^1 \frac{\partial M_2^T}{\partial x} + \frac{RT}{M_3^T}\phi_{33}^1 \frac{\partial M_3^T}{\partial x}\right)\right] \tag{18a}$$

$$V_mJ_2 = -\frac{M_2^T D_2^*}{RT}\left(\frac{RT}{M_2^T}\phi_{22}^1\frac{\partial M_2^T}{\partial x} + \frac{RT}{M_3^T}\phi_{23}^1\frac{\partial M_3^T}{\partial x}\right) - \frac{\xi_{PT} M_2^T D_2^*}{RT}\Big[M_1^T D_1^*\left(\frac{RT}{M_2^T}\phi_{12}^1\frac{\partial M_2^T}{\partial x} + \frac{RT}{M_3^T}\phi_{13}^1\frac{\partial M_3^T}{\partial x}\right) + M_2^T D_2^*\left(\frac{RT}{M_2^T}\phi_{22}^1\frac{\partial M_2^T}{\partial x} + \frac{RT}{M_3^T}\phi_{23}^1\frac{\partial M_3^T}{\partial x}\right) + M_3^T D_3^*\left(\frac{RT}{M_2^T}\phi_{32}^1\frac{\partial M_2^T}{\partial x} + \frac{RT}{M_3^T}\phi_{33}^1\frac{\partial M_3^T}{\partial x}\right)\Big] \quad (18b)$$

$$V_mJ_3 = -\frac{M_3^T D_3^*}{RT}\left(\frac{RT}{M_2^T}\phi_{32}^1\frac{\partial M_2^T}{\partial x} + \frac{RT}{M_3^T}\phi_{33}^1\frac{\partial M_3^T}{\partial x}\right) - \frac{\xi_{PT} M_3^T D_3^*}{RT}\Big[M_1^T D_1^*\left(\frac{RT}{M_2^T}\phi_{12}^1\frac{\partial M_2^T}{\partial x} + \frac{RT}{M_3^T}\phi_{13}^1\frac{\partial M_3^T}{\partial x}\right) + M_2^T D_2^*\left(\frac{RT}{M_2^T}\phi_{22}^1\frac{\partial M_2^T}{\partial x} + \frac{RT}{M_3^T}\phi_{23}^1\frac{\partial M_3^T}{\partial x}\right) + M_3^T D_3^*\left(\frac{RT}{M_2^T}\phi_{32}^1\frac{\partial M_2^T}{\partial x} + \frac{RT}{M_3^T}\phi_{33}^1\frac{\partial M_3^T}{\partial x}\right)\Big] \quad (18c)$$

Rearranging Equation 17 with respect to $\frac{\partial M_2^T}{\partial x}$ and $\frac{\partial M_3^T}{\partial x}$, we have

$$V_mJ_i = -\frac{M_i^T}{M_2^T}D_i^*[\phi_{i2}^1 + \xi_{PT}(M_1^T D_1^*\phi_{12}^1 + M_2^T D_2^*\phi_{22}^1 + M_3^T D_3^*\phi_{32}^1)]\frac{\partial M_2^T}{\partial x} - \frac{M_i^T}{M_3^T}D_i^*[\phi_{i3}^1 + \xi_{PT}(M_1^T D_1^*\phi_{13}^1 + M_2^T D_2^*\phi_{23}^1 + M_3^T D_3^*\phi_{33}^1)]\frac{\partial M_3^T}{\partial x} \quad (19)$$

Therefore, Equation 18 following the similar arrangements for different components can be expressed as

$$V_mJ_1 = -\frac{M_1^T}{M_2^T}D_1^*[\phi_{12}^1 + \xi_{PT}(M_1^T D_1^*\phi_{12}^1 + M_2^T D_2^*\phi_{22}^1 + M_3^T D_3^*\phi_{32}^1)]\frac{\partial M_2^T}{\partial x} - \frac{M_1^T}{M_3^T}D_1^*[\phi_{13}^1 + \xi_{PT}(M_1^T D_1^*\phi_{13}^1 + M_2^T D_2^*\phi_{23}^1 + M_3^T D_3^*\phi_{33}^1)]\frac{\partial M_3^T}{\partial x} \quad (20a)$$

$$V_mJ_2 = -\frac{M_2^T}{M_2^T}D_2^*[\phi_{22}^1 + \xi_{PT}(M_1^T D_1^*\phi_{12}^1 + M_2^T D_2^*\phi_{22}^1 + M_3^T D_3^*\phi_{32}^1)]\frac{\partial M_2^T}{\partial x} - \frac{M_2^T}{M_3^T}D_2^*[\phi_{23}^1 + \xi_{PT}(M_1^T D_1^*\phi_{13}^1 + M_2^T D_2^*\phi_{23}^1 + M_3^T D_3^*\phi_{33}^1)]\frac{\partial M_3^T}{\partial x} \quad (20b)$$

$$V_mJ_3 = -\frac{M_3^T}{M_2^T}D_3^*[\phi_{32}^1 + \xi_{PT}(M_1^T D_1^*\phi_{12}^1 + M_2^T D_2^*\phi_{22}^1 + M_3^T D_3^*\phi_{32}^1)]\frac{\partial M_2^T}{\partial x} - \frac{M_3^T}{M_3^T}D_3^*[\phi_{33}^1 + \xi_{PT}(M_1^T D_1^*\phi_{13}^1 + M_2^T D_2^*\phi_{23}^1 + M_3^T D_3^*\phi_{33}^1)]\frac{\partial M_3^T}{\partial x} \quad (20c)$$

Equating Equation 5a with 19, we have

$$D_{i2}^1 = \frac{M_i^T}{M_2^T}D_i^*\left[\phi_{i2}^1 + \xi_{PT}(M_1^T D_1^*\phi_{12}^1 + M_2^T D_2^*\phi_{22}^1 + M_3^T D_3^*\phi_{32}^1)\right] = \frac{M_i^T}{M_2^T}D_i^*\phi_{i2}^1(1 + W_{i2}) \quad (21a)$$

$$D_{i3}^1 = \frac{M_i^T}{M_3^T}D_i^*\left[\phi_{i3}^1 + \xi_{PT}(M_1^T D_1^*\phi_{13}^1 + M_2^T D_2^*\phi_{23}^1 + M_3^T D_3^*\phi_{33}^1)\right] = \frac{M_i^T}{M_3^T}D_i^*\phi_{i3}^1(1 + W_{i3}) \quad (21b)$$

Such that, the contribution of the vacancy wind effects can be expressed as

$$1 + W_{i2} = 1 + \xi_{PT}\left(M_1^T D_1^*\frac{\phi_{12}^1}{\phi_{i2}^1} + M_1^T D_2^*\frac{\phi_{22}^1}{\phi_{i2}^1} + M_3^T D_3^*\frac{\phi_{32}^1}{\phi_{i2}^1}\right) = 1 + \frac{2\left(M_1^T D_1^*\frac{\phi_{12}^1}{\phi_{i2}^1} + M_2^T D_2^*\frac{\phi_{22}^1}{\phi_{i2}^1} + M_3^T D_3^*\frac{\phi_{32}^1}{\phi_{i2}^1}\right)}{S_o(M_1^T D_1^* + M_2^T D_2^* + M_3^T D_3^*)} \quad (22a)$$

$$1 + W_{i3} = 1 + \xi_{PT}\left(M_1^T D_1^*\frac{\phi_{13}^1}{\phi_{i3}^1} + M_2^T D_2^*\frac{\phi_{23}^1}{\phi_{i3}^1} + M_3^T D_3^*\frac{\phi_{33}^1}{\phi_{i3}^1}\right) = 1 + \frac{2\left(M_1^T D_1^*\frac{\phi_{13}^1}{\phi_{i3}^1} + M_2^T D_2^*\frac{\phi_{23}^1}{\phi_{i3}^1} + M_3^T D_3^*\frac{\phi_{33}^1}{\phi_{i3}^1}\right)}{S_o(M_1^T D_1^* + M_2^T D_2^* + M_3^T D_3^*)} \quad (22b)$$

Since $\xi_{PT} = \frac{2}{S_o(M_1^T D_1^* + M_2^T D_2^* + M_3^T D_3^*)}$ and $S_o = 7.15$ in FCC crystal [32].

Therefore, one can see that Equation 21 reduces to Equation 13 when we neglect the vacancy wind effect by neglecting the terms $W_{i2}$ and $W_{i3}$.

Therefore, comparing Equations 5b-d and 21a-b, the intrinsic diffusion coefficients of individual components can be expressed as

$$D_{12}^1 = \frac{M_1^T}{M_2^T} D_1^* [\phi_{12}^1 + \xi_{PT}(M_1^T D_1^* \phi_{12}^1 + M_2^T D_2^* \phi_{22}^1 + M_3^T D_3^* \phi_{32}^1)] = \frac{M_1^T}{M_2^T} D_1^* \phi_{12}^1 (1 + W_{12}) \qquad (23a)$$

$$D_{13}^1 = \frac{M_1^T}{M_3^T} D_1^* [\phi_{13}^1 + \xi_{PT}(M_1^T D_1^* \phi_{13}^1 + M_2^T D_2^* \phi_{23}^1 + M_3^T D_3^* \phi_{33}^1)] = \frac{M_1^T}{M_3^T} D_1^* \phi_{13}^1 (1 + W_{13}) \qquad (23b)$$

$$D_{22}^1 = D_2^* [\phi_{22}^1 + \xi_{PT}(M_1^T D_1^* \phi_{12}^1 + M_2^T D_2^* \phi_{22}^1 + M_3^T D_3^* \phi_{32}^1)] = D_2^* \phi_{22}^1 (1 + W_{22}) \qquad (23c)$$

$$D_{23}^1 = \frac{M_2^T}{M_3^T} D_2^* [\phi_{23}^1 + \xi_{PT}(M_1^T D_1^* \phi_{13}^1 + M_2^T D_2^* \phi_{23}^1 + M_3^T D_3^* \phi_{33}^1)] = \frac{M_2^T}{M_3^T} D_2^* \phi_{23}^1 (1 + W_{23}) \qquad (23d)$$

$$D_{32}^1 = \frac{M_3^T}{M_2^T} D_3^* [\phi_{32}^1 + \xi_{PT}(M_1^T D_1^* \phi_{12}^1 + M_2^T D_2^* \phi_{22}^1 + M_3^T D_3^* \phi_{32}^1)] = \frac{M_3^T}{M_2^T} D_3^* \phi_{32}^1 (1 + W_{32}) \qquad (23e)$$

$$D_{33}^1 = D_3^* [\phi_{33}^1 + \xi_{PT}(M_1^T D_1^* \phi_{13}^1 + M_2^T D_2^* \phi_{23}^1 + M_3^T D_3^* \phi_{33}^1)] = D_3^* \phi_{33}^1 (1 + W_{33}) \qquad (23f)$$

One can then write the interdiffusion coefficients with the vacancy wind effect by replacing Equation 23 in Equation 6 to follow a similar step to calculate the tracer diffusion coefficients considering the vacancy wind effect.

### 2.2 Relations between the interdiffusion coefficients considering different component as the dependent variables in PT diffusion couples

Another important aspect should be described here which is not derived earlier in the PT diffusion couples. This is useful for the discussion in this article. If we consider the composition profiles of 2 and 3, we estimate the interdiffusion coefficients with component 1 as the dependent variable, which are described above. Similarly, we can estimate the interdiffusion coefficients with component 2 as the dependent variable when we estimate these data from the composition profiles of components 1 and 3. We can also estimate the interdiffusion coefficients considering component 3 as the dependent variable when we consider the composition profiles of components 1 and 2 for this estimation. Since we have $\tilde{J}_1 + \tilde{J}_2 + \tilde{J}_3 = 0$ and $\partial M_1^T + \partial M_2^T + \partial M_3^T = 0$, we can find how these interdiffusion coefficients considering different components as the dependent variable are related to one another.

The relations considering component 1 as the dependent variable are expressed in Equation 3. Similarly, the interdiffusion coefficients keeping component 2 as the dependent variable are expressed as

$$V_m \tilde{J}_1 = -\tilde{D}_{11}^2 \frac{\partial M_1^T}{\partial x} - \tilde{D}_{13}^2 \frac{\partial M_3^T}{\partial x} \qquad (24a)$$

$$V_m \tilde{J}_3 = -\tilde{D}_{31}^2 \frac{\partial M_1^T}{\partial x} - \tilde{D}_{33}^2 \frac{\partial M_3^T}{\partial x} \qquad (24b)$$

$$\tilde{J}_2 = -(\tilde{J}_1 + \tilde{J}_3) \qquad (24c)$$

The interdiffusion coefficients with component 3 as the dependent variable are expressed as

$$V_m \tilde{J}_1 = -\tilde{D}_{11}^3 \frac{\partial M_1^T}{\partial x} - \tilde{D}_{12}^3 \frac{\partial M_2^T}{\partial x} \qquad (25a)$$

$$V_m \tilde{J}_2 = -\tilde{D}_{21}^3 \frac{\partial M_1^T}{\partial x} - \tilde{D}_{22}^3 \frac{\partial M_2^T}{\partial x} \qquad (25b)$$

$$\tilde{J}_3 = -(\tilde{J}_1 + \tilde{J}_2) \qquad (25c)$$

Equating 24b and 3d, we have

$$-\tilde{D}_{31}^2 \frac{\partial M_1^T}{\partial x} - \tilde{D}_{33}^2 \frac{\partial M_3^T}{\partial x} = -\tilde{D}_{32}^1 \frac{\partial M_2^T}{\partial x} - \tilde{D}_{33}^1 \frac{\partial M_3^T}{\partial x}$$

Replacing, $\partial M_2^T = -(\partial M_1^T + \partial M_3^T)$

$$-\widetilde{D}_{31}^{2}\frac{\partial M_1^T}{\partial x}-\widetilde{D}_{33}^{2}\frac{\partial M_3^T}{\partial x} = -\widetilde{D}_{32}^{1}\left(-\frac{\partial M_1^T}{\partial x}-\frac{\partial M_3^T}{\partial x}\right)-\widetilde{D}_{33}^{1}\frac{\partial M_3^T}{\partial x}$$

$$-\widetilde{D}_{31}^{2}\frac{\partial M_1^T}{\partial x}-\widetilde{D}_{33}^{2}\frac{\partial M_3^T}{\partial x} = -\left(-\widetilde{D}_{32}^{1}\right)\frac{\partial M_1^T}{\partial x}-\left(\widetilde{D}_{33}^{1}-\widetilde{D}_{32}^{1}\right)\frac{\partial M_3^T}{\partial x}$$

Comparing the modified composition on both sides, we have

$$\widetilde{D}_{31}^{2} = -\widetilde{D}_{32}^{1} \tag{26a}$$

$$\widetilde{D}_{33}^{2} = \widetilde{D}_{33}^{1}-\widetilde{D}_{32}^{1} \tag{26b}$$

Equating 25b and 3c

$$-\widetilde{D}_{21}^{3}\frac{\partial M_1^T}{\partial x}-\widetilde{D}_{22}^{3}\frac{\partial M_2^T}{\partial x} = -\widetilde{D}_{22}^{1}\frac{\partial M_2^T}{\partial x}-\widetilde{D}_{23}^{1}\frac{\partial M_3^T}{\partial x}$$

Replacing $\partial M_3 = -(\partial M_1 + \partial M_2)$, we have

$$-\widetilde{D}_{21}^{3}\frac{\partial M_1^T}{\partial x}-\widetilde{D}_{22}^{3}\frac{\partial M_2^T}{\partial x} = -\widetilde{D}_{22}^{1}\frac{\partial M_2^T}{\partial x}-\widetilde{D}_{23}^{1}\left(-\frac{\partial M_1^T}{\partial x}-\frac{\partial M_2^T}{\partial x}\right)$$

$$-\widetilde{D}_{21}^{3}\frac{\partial M_1^T}{\partial x}-\widetilde{D}_{22}^{3}\frac{\partial M_2^T}{\partial x} = -\left(-\widetilde{D}_{23}^{1}\right)\frac{\partial M_1^T}{\partial x}-\left(\widetilde{D}_{22}^{1}-\widetilde{D}_{23}^{1}\right)\frac{\partial M_2^T}{\partial x}$$

Comparing the modified composition gradients on both sides, we have

$$\widetilde{D}_{21}^{3} = -\widetilde{D}_{23}^{1} \tag{26c}$$

$$\widetilde{D}_{22}^{3} = \widetilde{D}_{22}^{1}-\widetilde{D}_{23}^{1} \tag{26d}$$

In a PT couple, we can write

$$\tilde{J}_1 = -(\tilde{J}_2+\tilde{J}_3)$$

Expressing $\tilde{J}_1$ with Equation 24a, $\tilde{J}_2$ with Equation 3c, $\tilde{J}_3$ with Equation 3d, we have

$$-\widetilde{D}_{11}^{2}\frac{\partial M_1^T}{\partial x}-\widetilde{D}_{13}^{2}\frac{\partial M_3^T}{\partial x} = -\left[\left(-\widetilde{D}_{22}^{1}\frac{\partial M_2^T}{\partial x}-\widetilde{D}_{23}^{1}\frac{\partial M_3^T}{\partial x}\right)+\left(-\widetilde{D}_{32}^{1}\frac{\partial M_2^T}{\partial x}-\widetilde{D}_{33}^{1}\frac{\partial M_3^T}{\partial x}\right)\right]$$

Replacing $\partial M_1 = -(\partial M_2 + \partial M_3)$, we have

$$\widetilde{D}_{11}^{2}\frac{\partial M_2^T}{\partial x}-\left(\widetilde{D}_{13}^{2}-\widetilde{D}_{11}^{2}\right)\frac{\partial M_3^T}{\partial x} = \left(\widetilde{D}_{22}^{1}+\widetilde{D}_{32}^{1}\right)\frac{\partial M_2^T}{\partial x}+\left(\widetilde{D}_{23}^{1}+\widetilde{D}_{33}^{1}\right)\frac{\partial M_3^T}{\partial x}$$

Comparing the modified composition gradients on both sides, we have

$$\widetilde{D}_{11}^{2} = \widetilde{D}_{22}^{1}+\widetilde{D}_{32}^{1} \tag{26e}$$

$$\widetilde{D}_{13}^{2} = \widetilde{D}_{11}^{2}-\widetilde{D}_{23}^{1}-\widetilde{D}_{33}^{1} = \widetilde{D}_{22}^{1}+\widetilde{D}_{32}^{1}-\widetilde{D}_{23}^{1}-\widetilde{D}_{33}^{1} \tag{26f}$$

Final Expression of Equation 26f is found by replacing Equation 26e in it.

Again expressing $\tilde{J}_1$ with Equation 25a, $\tilde{J}_2$ with Equation 3c and $\tilde{J}_3$ with Equation 3d, we have

$$\tilde{J}_1 = -(\tilde{J}_2+\tilde{J}_3)$$

$$-\widetilde{D}_{11}^{3}\frac{\partial M_1^T}{\partial x}-\widetilde{D}_{12}^{3}\frac{\partial M_2^T}{\partial x} = -\left[-\left(\widetilde{D}_{22}^{1}+\widetilde{D}_{32}^{1}\right)\frac{\partial M_2^T}{\partial x}-\left(\widetilde{D}_{23}^{1}+\widetilde{D}_{33}^{1}\right)\frac{\partial M_3^T}{\partial x}\right]$$

Replacing $\partial M_1 = -(\partial M_2 + \partial M_3)$, we have

$$\left(\widetilde{D}_{11}^{3}-\widetilde{D}_{12}^{3}\right)\frac{\partial M_2^T}{\partial x}+\widetilde{D}_{11}^{3}\frac{\partial M_3^T}{\partial x} = \left(\widetilde{D}_{22}^{1}+\widetilde{D}_{32}^{1}\right)\frac{\partial M_2^T}{\partial x}+\left(\widetilde{D}_{23}^{1}+\widetilde{D}_{33}^{1}\right)\frac{\partial M_3^T}{\partial x}$$

Comparing both the sides, we have

$$\tilde{D}^3_{11} = \tilde{D}^1_{23} + \tilde{D}^1_{33} \tag{26g}$$

$$\tilde{D}^3_{12} = \tilde{D}^3_{11} - \tilde{D}^1_{22} - \tilde{D}^1_{32} = \tilde{D}^1_{23} + \tilde{D}^1_{33} - \tilde{D}^1_{22} - \tilde{D}^1_{32} \tag{26h}$$

Final Expression of Equation 26h is found by replacing Equation 26g in it.

Table 1 The relations between interdiffusion coefficients considering different components as the dependent variable with the same with component 1 as the dependent variables in different types of couples.

| | Quaternary diffusion couple | | Ternary/PT diffusion couple** | | PB diffusion couple | |
|---|---|---|---|---|---|---|
| CDV 2* | $\tilde{J}_1$ | $\tilde{D}^2_{11} = \tilde{D}^1_{22} + \tilde{D}^1_{32} + \tilde{D}^1_{42}$ | $\tilde{J}_1$ | $\tilde{D}^2_{11} = \tilde{D}^1_{22} + \tilde{D}^1_{32}$ | $\tilde{J}_1=\tilde{J}_2$ | $\tilde{D}^2_{11} = \tilde{D}^1_{22} (= \tilde{D})$ |
| | | $\tilde{D}^2_{13} = \tilde{D}^1_{22} + \tilde{D}^1_{32} + \tilde{D}^1_{42} - \tilde{D}^1_{23} - \tilde{D}^1_{33} - \tilde{D}^1_{43}$ | | $\tilde{D}^2_{13} = \tilde{D}^1_{22} + \tilde{D}^1_{32} - \tilde{D}^1_{23} - \tilde{D}^1_{33}$ | | |
| | | $\tilde{D}^2_{14} = \tilde{D}^1_{22} + \tilde{D}^1_{32} + \tilde{D}^1_{42} - \tilde{D}^1_{24} - \tilde{D}^1_{34} - \tilde{D}^1_{44}$ | | | | |
| | $\tilde{J}_3$ | $\tilde{D}^2_{31} = -\tilde{D}^1_{32}$ | $\tilde{J}_3$ | $\tilde{D}^2_{31} = -\tilde{D}^1_{32}$ | | |
| | | $\tilde{D}^2_{33} = \tilde{D}^1_{33} - \tilde{D}^1_{32}$ | | $\tilde{D}^2_{33} = \tilde{D}^1_{33} - \tilde{D}^1_{32}$ | | |
| | | $\tilde{D}^2_{34} = \tilde{D}^1_{34} - \tilde{D}^1_{32}$ | | | | |
| | $\tilde{J}_4$ | $\tilde{D}^2_{41} = -\tilde{D}^1_{42}$ | | | | |
| | | $\tilde{D}^2_{43} = \tilde{D}^1_{43} - \tilde{D}^1_{42}$ | | | | |
| | | $\tilde{D}^2_{44} = \tilde{D}^1_{44} - \tilde{D}^1_{42}$ | | | | |
| CDV 3* | $\tilde{J}_1$ | $\tilde{D}^3_{11} = \tilde{D}^1_{23} + \tilde{D}^1_{33} + \tilde{D}^1_{43}$ | $\tilde{J}_1$ | $\tilde{D}^3_{11} = \tilde{D}^1_{23} + \tilde{D}^1_{33}$ | | |
| | | $\tilde{D}^3_{12} = \tilde{D}^1_{23} + \tilde{D}^1_{33} + \tilde{D}^1_{43} - \tilde{D}^1_{22} - \tilde{D}^1_{32} - \tilde{D}^1_{42}$ | | $\tilde{D}^3_{12} = \tilde{D}^1_{23} + \tilde{D}^1_{33} - \tilde{D}^1_{22} - \tilde{D}^1_{32}$ | | |
| | | $\tilde{D}^3_{14} = \tilde{D}^1_{23} + \tilde{D}^1_{33} + \tilde{D}^1_{43} - \tilde{D}^1_{24} - \tilde{D}^1_{34} - \tilde{D}^1_{44}$ | | | | |
| | $\tilde{J}_2$ | $\tilde{D}^3_{21} = -\tilde{D}^1_{23}$ | $\tilde{J}_2$ | $\tilde{D}^3_{21} = -\tilde{D}^1_{23}$ | | |
| | | $\tilde{D}^3_{22} = \tilde{D}^1_{22} - \tilde{D}^1_{23}$ | | $\tilde{D}^3_{22} = \tilde{D}^1_{22} - \tilde{D}^1_{23}$ | | |
| | | $\tilde{D}^3_{24} = \tilde{D}^1_{24} - \tilde{D}^1_{23}$ | | | | |
| | $\tilde{J}_4$ | $\tilde{D}^3_{41} = -\tilde{D}^1_{43}$ | | | | |
| | | $\tilde{D}^3_{42} = \tilde{D}^1_{42} - \tilde{D}^1_{43}$ | | | | |
| | | $\tilde{D}^3_{44} = \tilde{D}^1_{44} - \tilde{D}^1_{43}$ | | | | |
| CDV 4* | $\tilde{J}_1$ | $\tilde{D}^4_{11} = \tilde{D}^1_{24} + \tilde{D}^1_{34} + \tilde{D}^1_{44}$ | | | | |
| | | $\tilde{D}^4_{12} = \tilde{D}^1_{24} + \tilde{D}^1_{34} + \tilde{D}^1_{44} - \tilde{D}^1_{22} - \tilde{D}^1_{32} - \tilde{D}^1_{42}$ | | | | |
| | | $\tilde{D}^4_{13} = \tilde{D}^1_{24} + \tilde{D}^1_{34} + \tilde{D}^1_{44} - \tilde{D}^1_{23} - \tilde{D}^1_{33} - \tilde{D}^1_{43}$ | | | | |
| | $\tilde{J}_2$ | $\tilde{D}^4_{21} = -\tilde{D}^1_{24}$ | | | | |
| | | $\tilde{D}^4_{22} = \tilde{D}^1_{22} - \tilde{D}^1_{24}$ | | | | |
| | | $\tilde{D}^4_{23} = \tilde{D}^1_{23} - \tilde{D}^1_{24}$ | | | | |
| | $\tilde{J}_3$ | $\tilde{D}^4_{31} = -\tilde{D}^1_{34}$ | | | | |
| | | $\tilde{D}^4_{32} = \tilde{D}^1_{32} - \tilde{D}^1_{34}$ | | | | |
| | | $\tilde{D}^4_{33} = \tilde{D}^1_{33} - \tilde{D}^1_{34}$ | | | | |

*Component dependent variable (CVD). **The relations in a PT diffusion couples are similar to the ternary system but the number of components in the system are different. Therefore, the values of interdiffusion coefficients are different and not comparable in systems with different number of components.

It should be noted here that these relations in PT couples are similar to the equations one would derive in a ternary system [23]. Let us now compare these relations with the relations of the interdiffusion coefficients, when estimated following the conventional method by developing the diffusion profiles of all the four components (if it would be possible by intersecting three diffusion couples) at the same composition (see supplementary files) since we have analyzed the data in NiCoFeCr system. These are listed in Table 1. It should be noted here that in a conventional method four-component system, we would estimate 9 interdiffusion coefficients (see supplementary file). Since one component is kept constant in the same four-component PT diffusion couples, we estimate 4 interdiffusion coefficients. Further, since two components are kept constant in a PB couple (discussed next) in the same four-component system, we estimate only one interdiffusion coefficient. It can be seen that the relations $\widetilde{D}_{31}^{2} = -\widetilde{D}_{32}^{1}$ and $\widetilde{D}_{21}^{3} = -\widetilde{D}_{23}^{1}$ are the same in both conventional and PT couples. However, these values will be different when estimated from different types of couples since these are related differently with the intrinsic diffusion coefficients (Equation 6a and supplementary file). This will be discussed further based on the data calculated in the NiCoFeCr system.

2.3 **Relations between the diffusion coefficients in a PB couple**

In this article, we also compare the data estimated in PB and PT couples in the same system to explain the similarities/differences between the intrinsic and interdiffusion coefficients. Let us assume that only components 1 and 2 develop the diffusion profiles keeping all the components constant throughout the interdiffusion zone. Therefore, we have

$$N_1 + N_2 = 1 - N_3 - N_4 \ldots - N_n = 1 - N_f = N_v \quad (27a)$$

$$\frac{N_1}{N_v} + \frac{N_2}{N_v} = 1 \quad (27b)$$

$$M_1^B + M_2^B = 1 \quad (27c)$$

where $N_f = N_3 + N_4 \ldots + N_n$ is the sum of the compositions of components which remain constant and $N_v = N_1 + N_2$ is the sum of compositions of components which contribute to the development of the diffusion profiles. $M_i^B$ is the modified composition profile in PB diffusion couple. It should be noted here that $M_i^B$ in PB and $M_i^T$ in PT diffusion couples are different because of the difference in $N_v$ in these couples. In a PB couple:, $\tilde{J}_1 + \tilde{J}_2 = 0$, $\tilde{J}_3 = \cdots = \tilde{J}_n = 0$, $\frac{dN_3}{dx} = \cdots = \frac{dN_n}{dx} = 0$ and, therefore, Equation 1 with respect to the modified composition profile for component 2 reduces to

$$V_m \tilde{J}_2 = -\widetilde{D}_{22}^{1} \frac{\partial M_2^B}{\partial x} = -\widetilde{D} \frac{\partial M_2^B}{\partial x} \quad (28a)$$

Similarly, when the interdiffusion coefficients are expressed for component 1 reduces to

$$V_m \tilde{J}_1 = -\widetilde{D}_{11}^{2} \frac{\partial M_1^B}{\partial x} = -\widetilde{D} \frac{\partial M_1^B}{\partial x} \quad (28b)$$

In a PB couple, we have $\partial M_1^B + \partial M_2^B = 0$ and $\tilde{J}_2 + \tilde{J}_1 = 0$. Therefore, we have $\widetilde{D}_{22}^{1} = \widetilde{D}_{11}^{2} = \widetilde{D}$. It further means that, like the binary system, we will have the same value of the interdiffusion coefficient estimated considering diffusion profile of any of the component for the estimation of the data in a PB couple. Because of this reason, we can estimate composition-dependent diffusion coefficients from a single PB diffusion couple similar to the binary system. Therefore, the interdiffusion flux and the diffusion profiles are also different in the PB diffusion couple compared to PT and conventional couples.

The interdiffusion flux in a PB couple can be expressed as (see supplementary file)

$$V_m \tilde{J}_i = -\frac{M_i^{B+} - M_i^{B-}}{2t} \left[ \left(1 - Y_{M_{i,B}}^{*}\right) \int_{x^{-\infty}}^{x^*} Y_{M_{i,B}} dx + Y_{M_{i,B}}^{*} \int_{x^*}^{x^{+\infty}} (1 - Y_{M_{i,B}}) dx \right] \quad (29a)$$

where $Y_{M_{i,B}} = \frac{M_i^{B*} - M_i^{B-}}{M_i^{B+} - M_i^{B-}}$ is the modified composition normalized variable in a PB couple.

Equating, Equation 28 and 29a, the interdiffusion coefficient in a PB couple can be directly estimated from the composition profile utilizing

$$\widetilde{D} = \frac{1}{2t} \left( \frac{\partial x}{\partial Y_{M_{i,B}}} \right)^{*} \left[ \left(1 - Y_{M_{i,B}}^{*}\right) \int_{x^{-\infty}}^{x^*} Y_{M_{i,B}} dx + Y_{M_{i,B}}^{*} \int_{x^*}^{x^{+\infty}} (1 - Y_{M_{i,B}}) dx \right] \quad (29b)$$

In lattice fixed frame of reference, the intrinsic fluxes are related to the intrinsic diffusion coefficients with respect to modified composition profiles. Following the same argument as above, it can be expressed from Equation 1d with respect to the modified composition profiles as

$$J_2 = -D_{22}^1 \frac{\partial M_2^B}{\partial x} \tag{30a}$$

$$J_1 = -D_{11}^2 \frac{\partial M_1^B}{\partial x} \tag{30b}$$

The intrinsic fluxes of different components are related to the vacancy flux by

$$J_1 + J_2 + J_v = 0 \tag{31a}$$

where the Kirkendall marker velocity $v_K$ is related to the vacancy flux $J_v$ and the molar volume $V_M$ by

$$v_K = V_m J_v \tag{31b}$$

The interdiffusion and intrinsic fluxes following the Darken relation [37] in a PB couple are related by

$$\tilde{J}_i = J_i + v_K \frac{M_i^B}{V_m} \tag{31c}$$

Taking sum of these interdiffusion fluxes for two components and since $\tilde{J}_1 + \tilde{J}_2 = 0$ and $M_1^B + M_2^B = 1$, we have

$$v_K = -V_m (J_1 + J_2) \tag{32}$$

From Equations 31 and 32, we have

$$\tilde{J}_i = J_i - M_i^B (J_1 + J_2) \tag{33}$$

Writing it for any of component 1 or 2 (for example component 2), we have

$$\tilde{J}_2 = M_1^B J_2 - M_2^B J_1 \tag{34}$$

Replacing Equation 28 and 30 in 34 and since $\partial M_1^B + \partial M_2^B = 0$, we have

$$\widetilde{D} = M_1^B D_{22}^1 + M_2^B D_{11}^2 \tag{35}$$

Since only two components develop the diffusion profiles in a PB couple, Equation 8 with respect to the modified compositions ($M_i^B$) can be expressed as

$$V_m J_i = -\frac{M_i^B D_i^*}{RT} \frac{\partial \mu_i}{dx} - \beta_{PB} M_i^B D_i^* V_m J_v \tag{36a}$$

where $\beta_{PB} = \frac{2}{(S_o+2)\sum_{j=1}^{2} M_j^B D_j^*}$ and $S_o$ is the structure factor.

Taking the sum of intrinsic fluxes of both the components and then replacing $J_v = -\sum_{j=1}^{2} J_j = -(J_1 + J_2)$, we have the relation for $J_v$. Replacing this in Equation 36a, we have

$$V_m J_i = -\frac{M_i^B D_i^*}{RT} \frac{\partial \mu_i}{dx} - \sum_{j=1}^{2} \frac{M_j^B D_j^*}{RT} \left(\xi_{PB} M_i^B D_i^*\right) \frac{\partial \mu_j}{dx} \tag{36b}$$

where $\xi_{PB} = \frac{\beta_{PB}}{1-\beta_{PB}\sum_{j=1}^{2} M_j^B D_j^*} = \frac{2}{S_o \sum_{j=1}^{2} M_j^B D_j^*}$

In a PB couple, component 1 and 2 develop the diffusion profiles keeping components 3 and 4 constant ($\frac{\partial N_3}{\partial x}$ *and* $\frac{\partial N_4}{\partial x} = 0$). Therefore, Equation 10b reduces to

$$\frac{\partial \mu_2}{\partial x} = \frac{\partial \mu_2}{\partial N_2} \frac{\partial N_2}{\partial x}$$

$$\frac{\partial \mu_2}{\partial x} = \frac{1}{N_2}\frac{RT\partial lna_2}{\partial lnN_2}\frac{\partial N_2}{\partial x}$$

$$\frac{\partial \mu_2}{\partial x} = \frac{RT}{N_2}\phi_{22}^1\frac{\partial N_2}{\partial x}$$

Dividing $N_v = N_1 + N_2$ in both numerator and denominator in the right-hand side of the equation, we have

$$\frac{\partial \mu_2}{\partial x} = \frac{RT}{M_2^B}\phi_{22}^1\frac{\partial M_2^B}{\partial x} \quad (37a)$$

Following the similar argument, the expression for the component 1 (i.e. by considering component 2 as the dependent variable) can be written as

$$\frac{\partial \mu_1}{\partial x} = \frac{RT}{M_1^B}\phi_{11}^2\frac{\partial M_1^B}{\partial x} \quad (37b)$$

If we neglect the vacancy wind effect (i.e. by neglecting all terms related to $\xi$), Equation 36b utilizing Equation 37 can be expressed as

$$V_m J_2 = -D_2^*\phi_{22}^1\frac{\partial M_2^B}{\partial x} \quad (38a)$$

The same can be written for the component 1 in the PB couple as

$$V_m J_1 = -D_1^*\phi_{11}^2\frac{\partial M_1^B}{\partial x} \quad (38b)$$

Comparing Equations 30 and 38, we have

$$D_{22}^1 = D_2^*\phi_{22}^1 \quad (39a)$$

$$D_{11}^2 = D_1^*\phi_{11}^2 \quad (39b)$$

Therefore, the interdiffusion and tracer diffusion coefficients are related by

$$\widetilde{D} = M_2^B D_1^*\phi_{11}^2 + M_1^B D_2^*\phi_{22}^1 \quad (39c)$$

It should be noted here that the tracer diffusion coefficients in PB couples can be estimated from the Kirkendall marker experiments. These are difficult or almost impossible to estimate at the intersection of PB couples (similar to the PT couples), which is discussed in the results and discussion section. If we consider the vacancy wind effect, Equation 36b can be expressed as

$$V_m J_1 = -\frac{M_1^B D_1^*}{RT}\frac{\partial \mu_1}{dx} - \frac{M_1^B D_1^*}{RT}(\xi_{PB} M_1^B D_1^*)\frac{\partial \mu_1}{dx} - \frac{M_2^B D_2^*}{RT}(\xi_{PB} M_1^B D_1^*)\frac{\partial \mu_2}{dx} \quad (40a)$$

$$V_m J_2 = -\frac{M_2^B D_2^*}{RT}\frac{\partial \mu_2}{dx} - \frac{M_1^B D_1^*}{RT}(\xi_{PB} M_2^B D_2^*)\frac{\partial \mu_1}{dx} - \frac{M_2^B D_2^*}{RT}(\xi_{PB} M_2^B D_2^*)\frac{\partial \mu_2}{dx} \quad (40b)$$

Replacing Equation 30 and 37 in 40, we have

$$D_{11}^2\frac{\partial M_1^B}{\partial x} = D_1^*\phi_{11}^2\frac{\partial M_1^B}{\partial x} + D_1^*\phi_{11}^2(\xi_{PB} M_1^B D_1^*)\frac{\partial M_1^B}{\partial x} + D_2^*\phi_{22}^1(\xi_{PB} M_1^B D_1^*)\frac{\partial M_2^B}{\partial x}$$

$$D_{22}^1\frac{\partial M_2^B}{\partial x} = D_2^*\phi_{22}^1\frac{\partial M_2^B}{\partial x} + D_1^*\phi_{11}^2(\xi_{PB} M_2^B D_2^*)\frac{\partial M_1^B}{\partial x} + D_2^*\phi_{22}^1(\xi_{PB} M_2^B D_2^*)\frac{\partial M_2^B}{\partial x}$$

These can be rearranged considering $\partial M_1^B + \partial M_2^B = 0$ as

$$D_{11}^2 = D_1^*[\phi_{11}^2 + M_1^B\xi_{PB}(D_1^*\phi_{11}^2 - D_2^*\phi_{22}^1)] \quad (41a)$$

$$D_{22}^1 = D_2^*[\phi_{22}^1 - M_2^B\xi_{PB}(D_1^*\phi_{11}^2 - D_2^*\phi_{22}^1)] \quad (41b)$$

These can be further rearranged using the relation for $\xi_{PB}$ $\left(=\frac{2}{S_o\sum_{j=1}^{2}M_j^B D_j^*}\right)$ as

$$D_{11}^2 = D_1^*\left[\phi_{11}^2 + \frac{2M_1^B(D_1^*\phi_{11}^2 - D_2^*\phi_{22}^1)}{S_o(M_1^B D_1^* + M_2^B D_2^*)}\right] = D_1^*\phi_{11}^2\left[1 + \frac{2M_1^B\left(D_1^* - D_2^*\frac{\phi_{22}^1}{\phi_{11}^2}\right)}{S_o(M_1^B D_1^* + M_2^B D_2^*)}\right] \quad (42a)$$

$$D_{22}^1 = D_2^*\left[\phi_{22}^1 - \frac{2M_2^B(D_1^*\phi_{11}^2 - D_2^*\phi_{22}^1)}{S_o(M_1^B D_1^* + M_2^B D_2^*)}\right] = D_2^*\phi_{22}^2\left[1 - \frac{2M_2^B\left(D_1^*\frac{\phi_{11}^2}{\phi_{22}^1} - D_2^*\right)}{S_o(M_1^B D_1^* + M_2^B D_2^*)}\right] \quad (42b)$$

Therefore, the contribution to the vacancy wind effect is expressed as $1 + W_1 = 1 + \frac{2M_1^B\left(D_1^* - D_2^*\frac{\phi_{22}^1}{\phi_{11}^2}\right)}{S_o(M_1^B D_1^* + M_2^B D_2^*)}$ and $1 - W_2 = 1 - \frac{2M_2\left(D_1^*\frac{\phi_{11}^2}{\phi_{22}^1} - D_2^*\right)}{S_o(M_1^B D_1^* + M_2^B D_2^*)}$

In many systems, we will have similar values of $\phi_{11}^2$ and $\phi_{22}^1$ and one can calculate the vacancy wind effect from $1 + W_1 = 1 + \frac{2M_1^B(D_1^* - D_2^*)}{S_o(M_1^B D_1^* + M_2^B D_2^*)}$ and $1 - W_2 = 1 - \frac{2M_2(D_1^* - D_2^*)}{S_o(M_1^B D_1^* + M_2^B D_2^*)}$ without introducing much error if the contribution from this effect is negligible [39].

The interdiffusion coefficients with respect to the vacancy wind by replacing Equation 41 in Equation 39c can be expressed as

$$\begin{aligned}
\widetilde{D} &= M_2^B D_1^*[\phi_{11}^2 + M_1^B\xi_{PB}(D_1^*\phi_{11}^2 - D_2^*\phi_{22}^1)] + M_1^B D_2^*[\phi_{22}^1 - M_2^B\xi_{PB}(D_1^*\phi_{11}^2 - D_2^*\phi_{22}^1)] \\
&= M_2^B D_1^*\phi_{11}^2 + M_1^B D_2^*\phi_{22}^1 + M_1^B M_2^B D_1^*\xi_{PB}[(D_1^*\phi_{11}^2 - D_2^*\phi_{22}^1)] - M_1^B M_2^B D_2^*\xi_{PB}[(D_1^*\phi_{11}^2 - D_2^*\phi_{22}^1)] \\
&= M_2^B D_1^*\phi_{11}^2 + M_1^B D_2^*\phi_{22}^1 + M_1^B M_2^B\xi_{PB}[D_1^*(D_1^*\phi_{11}^2 - D_2^*\phi_{22}^1) - D_2^*(D_1^*\phi_{11}^2 - D_2^*\phi_{22}^1)] \\
&= M_2^B D_1^*\phi_{11}^2 + M_1^B D_2^*\phi_{22}^1 + M_1^B M_2^B\xi_{PB}[D_1^*\phi_{11}^2(D_1^* - D_2^*) + D_2^*\phi_{22}^1(D_2^* - D_1^*)] \\
&= M_2^B D_1^*\phi_{11}^2 + M_1^B D_2^*\phi_{22}^1 + M_1^B M_2^B\xi_{PB}(D_1^* - D_2^*)(D_1^*\phi_{11}^2 - D_2^*\phi_{22}^1) \\
&= M_2^B D_1^*\phi_{11}^2 + M_1^B D_2^*\phi_{22}^1 + \frac{2M_1^B M_2^B}{S_o}\frac{(D_1^* - D_2^*)(D_1^*\phi_{11}^2 - D_2^*\phi_{22}^1)}{(M_1^B D_1^* + M_2^B D_2^*)} \\
&= (M_2^B D_1^*\phi_{11}^2 + M_1 D_2^*\phi_{22}^1)\left[1 + \frac{2M_1^B M_2^B(D_1^* - D_2^*)(D_1^*\phi_{11}^2 - D_2^*\phi_{22}^1)}{S_o(M_1^B D_1^* + M_2^B D_2^*)(M_2^B D_1^*\phi_{11}^2 + M_1^B D_2^*\phi_{22}^2)}\right] \quad (43)
\end{aligned}$$

Therefore $W_{AB} = 1 + \frac{2M_1^B M_2^B(D_1^* - D_2^*)(D_1^*\phi_{11}^2 - D_2^*\phi_{22}^1)}{S_o(M_1^B D_1^* + M_2^B D_2^*)(M_2^B D_1^*\phi_{11}^2 + M_1^B D_2^*\phi_{22}^2)}$

It should be noted here that in a binary system, these thermodynamic parameters are equal because of the Gibbs-Duhem equation $N_1 d\mu_1 + N_2 d\mu_2 = 0$ such that $\phi_{11}^2$ and $\phi_{22}^1$ transform to $\emptyset = \frac{\partial lna_1}{\partial lnN_1} = \frac{\partial lna_2}{\partial lnN_2}$ $(i = 1{,}2)$ and relations given in Equations 42 and 43 reduce to the equations proposed in a binary system by Manning [31].

## 3. Results and discussion

For utilization of the method proposed, we have considered the PT diffusion couples produced in the NiCoFeCr system which intersect at a composition close to the equiatomic composition, as shown in Figure 1. It can be seen that Cr is kept constant at the composition of 25.7 at.% in both the diffusion couples. This component remains constant even in the interdiffusion zone indicating the presence of ideal/near-ideal diffusion profiles. To facilitate a comparison with the data measured by the radiotracer method, a relatively small composition range was selected such that the diffusion couples intersect at a composition close to the equiatomic composition. The diffusion paths of the diffusion profiles cannot be predicted a priori without the knowledge of the composition-dependent diffusion matrix of the components and reliable thermodynamic parameters. The composition of intersection can be determined by plotting modified compositions on Gibb's triangle. By converting composition $N_i$ to $M_i^T$, we superimpose the PT Ni-Co-Fe (fixed Cr) profiles on a ternary Ni-Co-Fe Gibb's triangle. Once the modified composition at the intersection is found, the composition of the intersection can be calculated back from Equation 2. Following, the interdiffusion coefficients are estimated utilizing Equation 3. As listed in Table 2, a set of two interdiffusion fluxes are considered at a time for estimation of the interdiffusion coefficients with different components as the dependent variables (see supplementary file). For instance, when these are estimated considering the interdiffusion flux of components Co and Fe, we estimate the interdiffusion coefficients with Ni as the dependent variable. The data estimated with the different component as the dependent variable are found to be related between themselves following Equation 26.

Table 2 The interdiffusion coefficients estimated considering different components as the dependent variables estimated directly from the composition profiles at 1200°C at the composition of $N_{Ni}$ = 24.7, $N_{Co}$ = 24.9, $N_{Fe}$ = 24.7, $N_{Cr}$ = 25.7 at.%. The back-calculated interdiffusion coefficients from the average of tracer diffusion coefficients are compared by considering and neglecting the vacancy wind effect.

**Ni dependent variable**

| Type of estimation | $\tilde{D}^{Ni}_{CoCo}$ ($\times 10^{-15}$ m²/s) | $\tilde{D}^{Ni}_{CoFe}$ ($\times 10^{-15}$ m²/s) | $\tilde{D}^{Ni}_{FeFe}$ ($\times 10^{-15}$ m²/s) | $\tilde{D}^{Ni}_{FeCo}$ ($\times 10^{-15}$ m²/s) |
|---|---|---|---|---|
| Estimated directly from the diffusion profiles | 2.42±0.3 | -2.18±0.3 | 6.49±1.0 | -0.41±0.06 |
| Back-calculated from tracer diffusion coefficients considering VWE | 2.38±0.3 | -2.26±0.3 | 6.49±1.0 | -0.32±0.05 |
| Calculated neglecting VWE | 2.37±0.3 | -2.08±0.3 | 6.15±0.9 | -0.31±0.05 |

**Co dependent variable**

| Type of estimation | $\tilde{D}^{Co}_{FeFe}$ ($\times 10^{-15}$ m²/s) | $\tilde{D}^{Co}_{FeNi}$ ($\times 10^{-15}$ m²/s) | $\tilde{D}^{Co}_{NiNi}$ ($\times 10^{-15}$ m²/s) | $\tilde{D}^{Co}_{NiFe}$ ($\times 10^{-15}$ m²/s) |
|---|---|---|---|---|
| Estimated directly from the diffusion profiles | 6.90±0.9 | 0.41±0.06 | 2.01±0.3 | -2.30±0.3 |
| Back-calculated from tracer diffusion coefficients considering VWE | 6.81±0.9 | 0.32±0.05 | 2.06±0.3 | -2.17±0.3 |
| Calculated neglecting VWE | 6.46±0.9 | 0.31±0.05 | 2.06±0.3 | -2.01±0.3 |

**Fe dependent variable**

| Type of estimation | $\tilde{D}^{Fe}_{CoCo}$ ($\times 10^{-15}$ m²/s) | $\tilde{D}^{Fe}_{CoNi}$ ($\times 10^{-15}$ m²/s) | $\tilde{D}^{Fe}_{NiNi}$ ($\times 10^{-15}$ m²/s) | $\tilde{D}^{Fe}_{NiCo}$ ($\times 10^{-15}$ m²/s) |
|---|---|---|---|---|
| Estimated directly from the diffusion profiles | 4.60±0.7 | 2.18±0.3 | 4.31±0.6 | 2.30±0.3 |
| Back-calculated from tracer diffusion coefficients considering VWE | 4.64±0.7 | 2.26±0.3 | 4.23±0.6 | 2.17±0.3 |
| Calculated neglecting VWE | 4.45±0.8 | 2.08±0.3 | 4.07±0.6 | 2.01±0.3 |

For the calculation of the tracer diffusion coefficients, the thermodynamic factors at 1200°C are computed using Thermo-Calc (TCHEA2 database) at the composition of the intersection of the PT diffusion couples i.e. at $N_{Ni}$ = 24.7, $N_{Co}$ = 24.9, $N_{Fe}$ = 24.7, $N_{Cr}$ = 25.7 at.%. These are listed in Table 3. For the correctness of calculation of the tracer diffusion coefficients, we utilized the relations considering the vacancy wind effect. Therefore, first, the relations expressed in Equation 23 for intrinsic diffusion coefficients are replaced in Equation 6 for expressing the relations between interdiffusion and tracer diffusion coefficients. In these PT couples, Ni, Co and Fe produce the diffusion profiles. It means that we can calculate the tracer diffusion coefficients of these components only. Therefore, we need three independent equations to solve. Moreover, to understand the stability of the solution, these are estimated from the interdiffusion coefficients with the different component as the dependent variable. As listed in Table 4, we find different sets of solutions considering different combinations of the interdiffusion coefficients at a time. It can be seen the calculated values are very similar and within the acceptable range of errors. We have calculated the average tracer diffusion coefficients considering all solutions. To cross-check, we have calculated back the interdiffusion coefficients considering these average tracer diffusion coefficients to find an excellent match with the data estimated directly from the diffusion profile as shown in Table 2. Further, as described in [4], the tracer diffusion coefficients are estimated at the locations of the Kirkendall marker plane in PB diffusion couples at the compositions close to the composition of the intersection of PT couples. These are also listed in Table 5. There is a difference the tracer diffusion coefficients are calculated utilizing another thermodynamic database in Thermo-Calc with a minor difference. Even the tracer diffusion coefficients measured by radiotracer method is also available [40]. The values at 1200°C are extracted by extending the data measured in the range of 650-1100°C. An excellent match is found in the data estimated following different methods, which can be understood from Figure 2. The very minor difference in data following different methods might also be related to the minor difference in compositions at which these data are estimated. This indicates the reliability of this method for calculating the tracer diffusion coefficients following the newly proposed equations at the composition of the intersection of PT couples in a multicomponent system. It should be noted here that, we could estimate the tracer diffusion coefficients utilizing Equation 14 i.e. by neglecting the vacancy wind effect. We found the values as $D_{Ni}^* = 2.35 \pm 0.4 \times 10^{-15}$, $D_{Co}^* = 2.33 \pm 0.4 \times 10^{-15}$ and $D_{Fe}^* = 7.58 \pm 1.1 \times 10^{-15}$ m$^2$/s. It means that the error in calculation is very small for Ni and Co tracer diffusion coefficients when we neglect the vacancy wind effect, which are estimated as $D_{Ni}^* = 2.38 \pm 0.4 \times 10^{-15}$ and $D_{Co}^* = 2.37 \pm 0.4 \times 10^{-15}$ m$^2$/s. However, there is an error in the estimated Fe tracer diffusion coefficients by ~7.5%, which is found to be $D_{Fe}^*$ =7.05±1.1$\times 10^{-15}$ m$^2$/s when the vacancy wind effect is considered. We can calculate the PB interdiffusion coefficients to compare with the data estimated in PB Ni-Co(fixed Fe,Cr) and PB Fe-Cr (fixed Ni,Co) diffusion couples utilizing Equations 39c and the tracer diffusion coefficients estimated following the radiotracer and PT diffusion couple methods. The details of the PB diffusion couples can be found in [4]. Again, an excellent match is found, as shown in Figure 3. However, a little difference is noticed when interdiffusion coefficients are calculated utilizing the tracer diffusion coefficients estimated from polycrystalline samples following the same radiotracer method as reported by Vaidya et al. [41] although the difference is still within the range of experimental error.

Table 3 The thermodynamic factors considering different components as the dependent variables at $N_{Ni}$ = 24.7, $N_{Co}$ = 24.9, $N_{Fe}$ = 24.7, $N_{Cr}$ = 25.7 at.% at 1200°C extracted from Thermo-Calc (TCHEA2 database).

| Ni dependent variable | | | Co dependent variable | | | Fe dependent variable | | | Cr dependent variable | | |
|---|---|---|---|---|---|---|---|---|---|---|---|
| $\phi_{NiCo}^{Ni}$ | $\phi_{NiFe}^{Ni}$ | $\phi_{NiCr}^{Ni}$ | $\phi_{CoNi}^{Co}$ | $\phi_{CoFe}^{Co}$ | $\phi_{CoCr}^{Co}$ | $\phi_{FeNi}^{Fe}$ | $\phi_{FeCo}^{Fe}$ | $\phi_{FeCr}^{Fe}$ | $\phi_{CrNi}^{Cr}$ | $\phi_{CrCo}^{Cr}$ | $\phi_{CrFe}^{Cr}$ |
| -0.89 | -1.06 | -1.18 | -0.99 | -1.22 | -1.35 | -1.09 | -1.15 | -1.03 | -1.29 | -1.36 | -1.11 |
| $\phi_{CoCo}^{Ni}$ | $\phi_{CoFe}^{Ni}$ | $\phi_{CoCr}^{Ni}$ | $\phi_{NiNi}^{Co}$ | $\phi_{NiFe}^{Co}$ | $\phi_{NiCr}^{Co}$ | $\phi_{NiNi}^{Fe}$ | $\phi_{NiCo}^{Fe}$ | $\phi_{NiCr}^{Fe}$ | $\phi_{NiNi}^{Cr}$ | $\phi_{NiCo}^{Cr}$ | $\phi_{NiFe}^{Cr}$ |
| 0.99 | -0.22 | -0.32 | 0.89 | -0.17 | -0.26 | 1.06 | 0.17 | -0.08 | 1.14 | 0.25 | 0.07 |
| $\phi_{FeCo}^{Ni}$ | $\phi_{FeFe}^{Ni}$ | $\phi_{FeCr}^{Ni}$ | $\phi_{FeNi}^{Co}$ | $\phi_{FeFe}^{Co}$ | $\phi_{FeCr}^{Co}$ | $\phi_{CoNi}^{Fe}$ | $\phi_{CoCo}^{Fe}$ | $\phi_{CoCr}^{Fe}$ | $\phi_{CoNi}^{Cr}$ | $\phi_{CoCo}^{Cr}$ | $\phi_{CoFe}^{Cr}$ |
| -0.05 | 1.09 | 0.11 | 0.05 | 1.15 | 0.16 | 0.22 | 1.22 | -0.09 | 0.31 | 1.31 | 0.09 |
| $\phi_{CrCo}^{Ni}$ | $\phi_{CrFe}^{Ni}$ | $\phi_{CrCr}^{Ni}$ | $\phi_{CrNi}^{Co}$ | $\phi_{CrFe}^{Co}$ | $\phi_{CrCr}^{Co}$ | $\phi_{CrNi}^{Fe}$ | $\phi_{CrCo}^{Fe}$ | $\phi_{CrCr}^{Fe}$ | $\phi_{FeNi}^{Cr}$ | $\phi_{FeCo}^{Cr}$ | $\phi_{FeFe}^{Cr}$ |
| -0.06 | 0.18 | 1.34 | 0.06 | 0.24 | 1.40 | -0.18 | -0.24 | 1.15 | -0.11 | -0.16 | 0.99 |

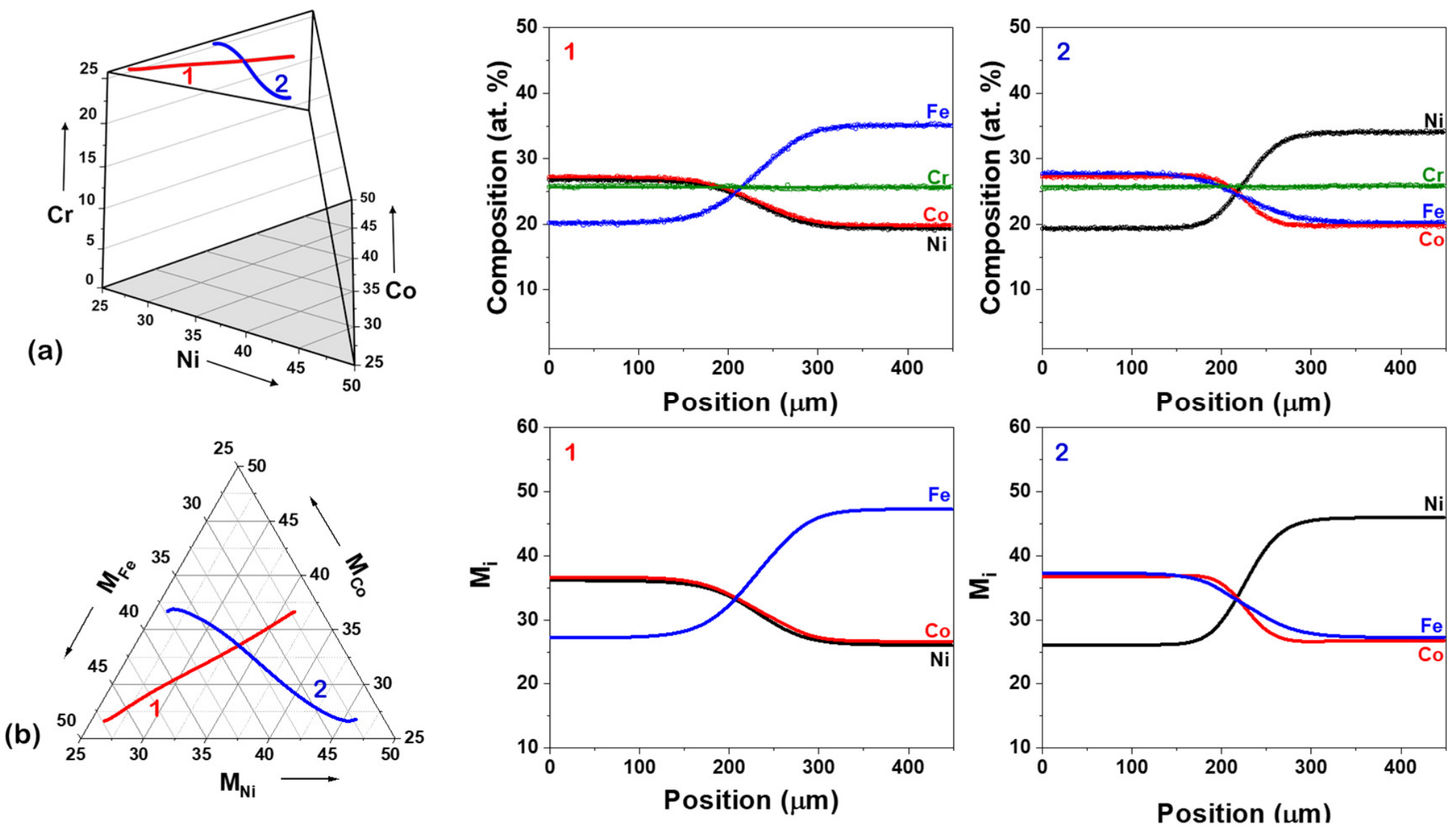

Figure 1 The pseudo-ternary (PT) diffusion couples annealed at 1200°C for 50h (a) Composition, (b) modified composition profiles and their diffusion paths on Gibb's triangle.

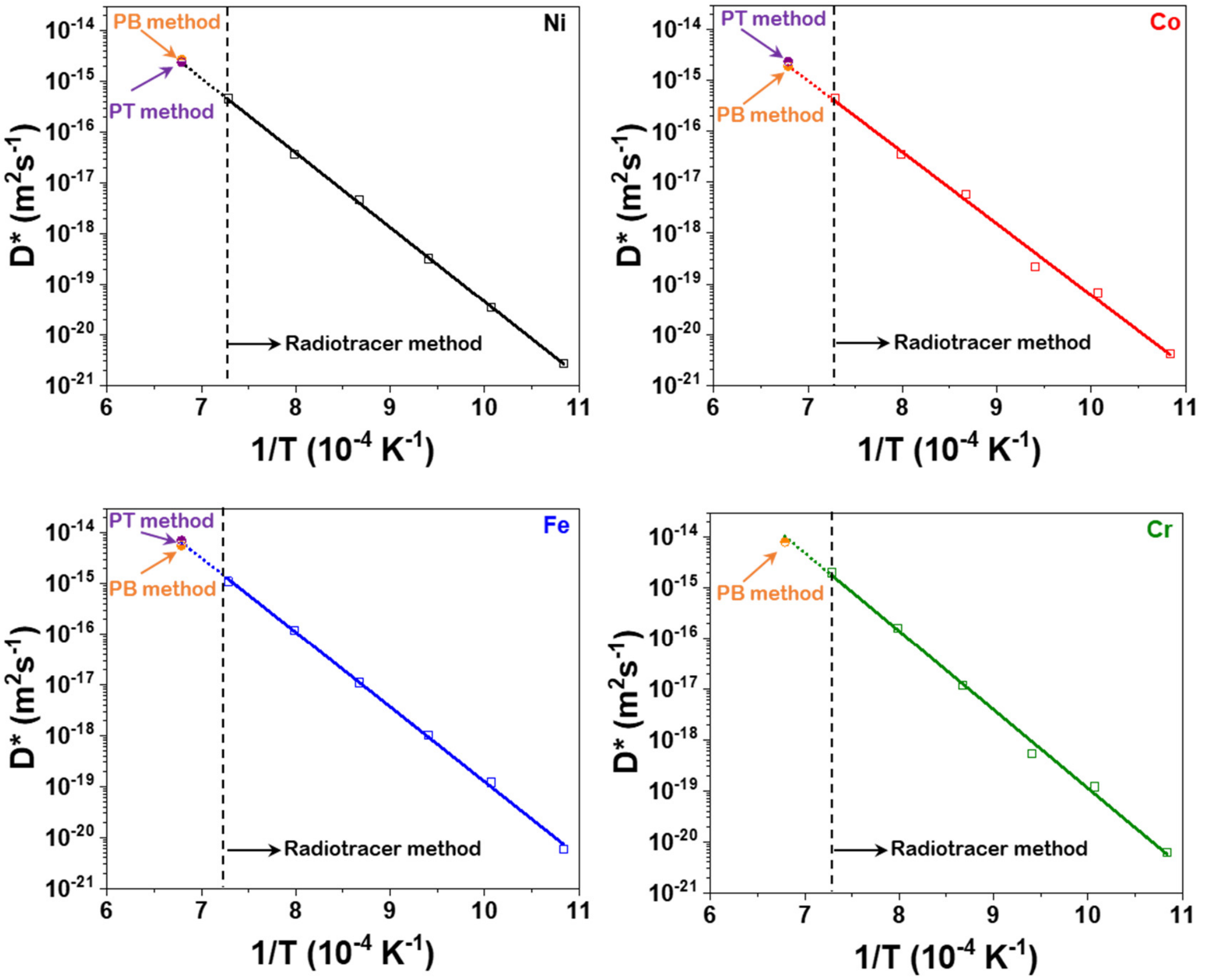

Figure 2 Comparison of the tracer diffusion coefficients estimated following the PT, PB diffusion couple methods and the direct measurement by the radiotracer method in single crystal alloy [40].

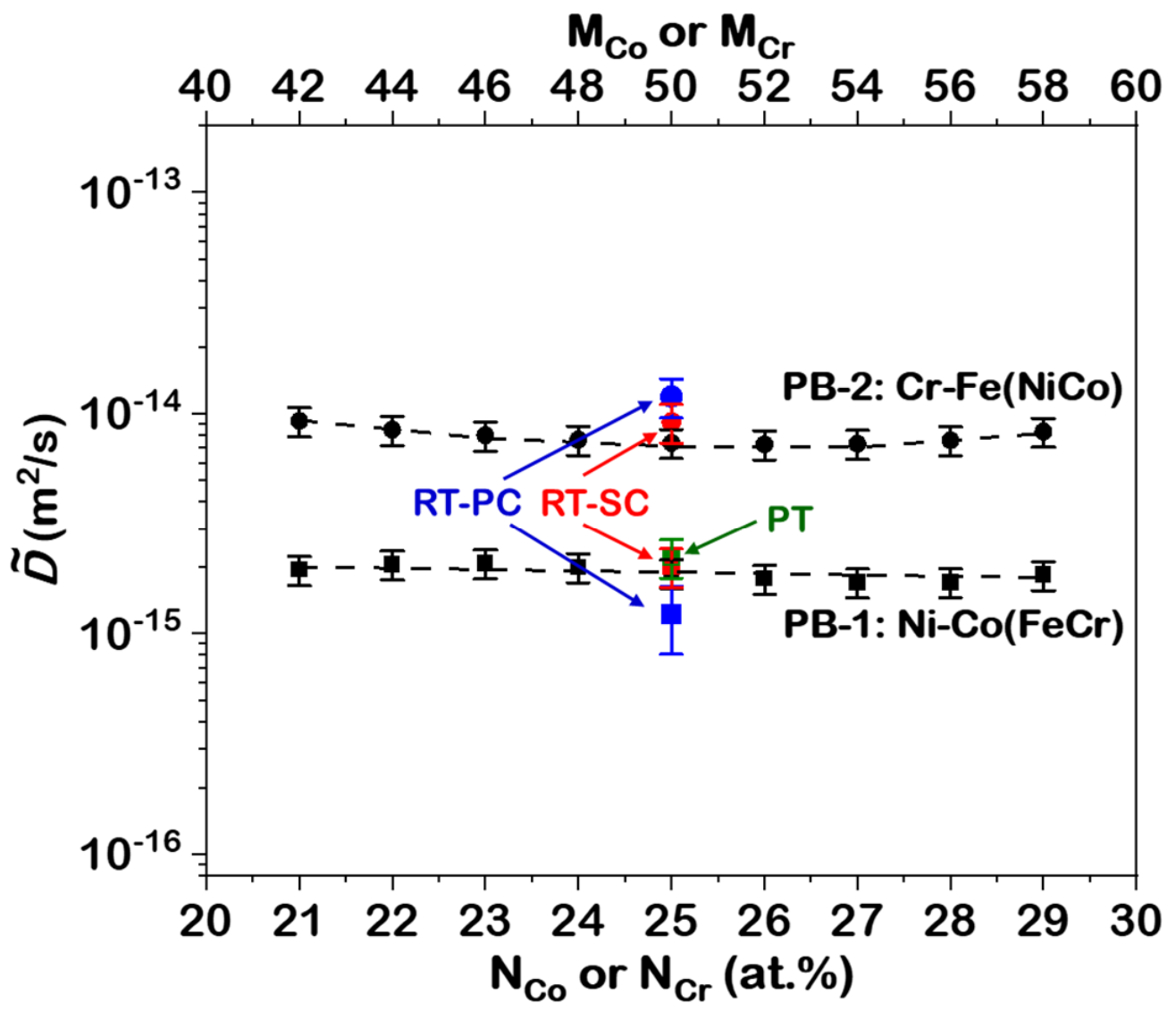

Figure 3 Comparison of the PB interdiffusion coefficients estimated directly from the PB diffusion couples and calculated from the tracer diffusion coefficients estimated following different methods (PT: PT diffusion couples, RT-SC: radiotracer method in single crystal alloy [40], RT-PC: radiotracer method in polycrystalline alloy [41].

Table 4 Calculated tracer diffusion coefficients from the estimated interdiffusion coefficients considering the vacancy wind effect (VWE) at $N_{Ni}$ = 24.7, $N_{Co}$ = 24.9, $N_{Fe}$ = 24.7, $N_{Cr}$ = 25.7 at.%. at 1200°C.

| | $D^*_{Ni}$ (×10$^{-15}$ m$^2$/s) | $D^*_{Co}$ (×10$^{-15}$ m$^2$/s) | $D^*_{Fe}$ (×10$^{-15}$ m$^2$/s) |
|---|---|---|---|
| $\tilde{D}^{Ni}_{CoCo}, \tilde{D}^{Ni}_{CoFe}, \tilde{D}^{Ni}_{FeCo}$ | 2.14 | 2.57 | 6.51 |
| $\tilde{D}^{Ni}_{CoCo}, \tilde{D}^{Ni}_{FeCo}, \tilde{D}^{Ni}_{FeFe}$ | 2.19 | 2.53 | 7.10 |
| $\tilde{D}^{Ni}_{CoFe}, \tilde{D}^{Ni}_{FeCo}, \tilde{D}^{Ni}_{FeFe}$ | 2.78 | 3.06 | 6.94 |
| $\tilde{D}^{Ni}_{CoFe}, \tilde{D}^{Ni}_{FeFe}, \tilde{D}^{Ni}_{CoCo}$ | 2.62 | 2.30 | 6.99 |
| $\tilde{D}^{Co}_{NiNi}, \tilde{D}^{Co}_{NiFe}, \tilde{D}^{Co}_{FeNi}$ | 2.19 | 2.47 | 7.39 |
| $\tilde{D}^{Co}_{NiNi}, \tilde{D}^{Co}_{FeNi}, \tilde{D}^{Co}_{FeFe}$ | 2.16 | 2.50 | 7.04 |
| $\tilde{D}^{Co}_{NiFe}, \tilde{D}^{Co}_{FeNi}, \tilde{D}^{Co}_{FeFe}$ | 1.79 | 2.16 | 7.11 |
| $\tilde{D}^{Co}_{NiFe}, \tilde{D}^{Co}_{FeFe}, \tilde{D}^{Co}_{NiNi}$ | 2.36 | 2.21 | 7.12 |
| $\tilde{D}^{Fe}_{NiNi}, \tilde{D}^{Fe}_{NiCo}, \tilde{D}^{Fe}_{CoNi}$ | 2.53 | 1.93 | 6.98 |
| $\tilde{D}^{Fe}_{NiNi}, \tilde{D}^{Fe}_{NiCo}, \tilde{D}^{Fe}_{CoCo}$ | 2.37 | 2.18 | 7.22 |
| $\tilde{D}^{Fe}_{NiCo}, \tilde{D}^{Fe}_{CoNi}, \tilde{D}^{Fe}_{CoCo}$ | 2.86 | 2.20 | 7.23 |
| $\tilde{D}^{Fe}_{CoNi}, \tilde{D}^{Fe}_{CoCo}, \tilde{D}^{Fe}_{NiNi}$ | 2.59 | 2.37 | 6.95 |
| **Average** | **2.38±0.4** | **2.37±0.4** | **7.05±1** |

Following, we have estimated the intrinsic diffusion coefficients following Equations 13 and 23 i.e. by neglecting and considering the role of vacancy wind effects, which are listed in Table 6. A direct correlation between the intrinsic and thermodynamic factors can be noticed. The positive or negative values of the intrinsic diffusion coefficients indicating the diffusional interactions between the components are directly related to the sign of the thermodynamic factors. Comparison of the data indicates the role of vacancy wind effect on different intrinsic diffusion coefficients, which are listed in Table 7. It can be seen that a few intrinsic diffusion coefficients are strongly affected by this. The contribution of this in the PB couples are also listed. This indicates that vacancy wind effect is negligible in both the PB couples considered in this study. However, this cannot be neglected in a multicomponent diffusion. To examine the role of this effect in the interdiffusion coefficients, we have estimated these parameters by replacing the values calculated by Equation 13 in Equation 6. The interdiffusion coefficients are affected maximum by ~9%, which is much lower than the influence on few intrinsic diffusion coefficients. Interdiffusion coefficients are kind of average of a set of certain intrinsic diffusion coefficients and, therefore, do not reflect how the intrinsic diffusion coefficients are affected individually. Moreover, as expected the intrinsic diffusion coefficients estimated from PB and PT couple are comparable within the range of experimental error, which should be similar if the contribution from vacancy wind effect is negligible which is different in different types of couples. Therefore, we can estimate both tracer and intrinsic diffusion coefficients from the multicomponent diffusion profiles by producing PT couples, which would not be possible otherwise in conventional method by intersecting the diffusion couples.

The diffusion coefficients should be estimated from only ideal or near-ideal diffusion profiles in which the non-ideality or the presence of hump is not found much beyond the standard range of error of composition measurements by WDS in EPMA. Otherwise, the error in calculation could be higher. Illogical values might be calculated if these are estimated from poorly prepared diffusion couples or if the thermodynamic details are not reliable.

As already mentioned, It can be seen from the relations between the intrinsic and tracer diffusion coefficients in PB (Equation 39c), PT (Equation 13) and conventional (see supplementary file) diffusion couples that the main and cross intrinsic diffusion coefficients of a component are the same when the vacancy wind effect is negligible irrespective of the type of diffusion couples used for the estimation of the data. However, there is a difference in the number of intrinsic diffusion coefficients involved in different types of couples. All the twelve independent intrinsic diffusion coefficients in a four-component system contribute to the interdiffusion coefficients and, therefore, on the development of the diffusion profiles in a conventional diffusion couple in which all the components are made to diffuse. When an ideal PT profile is developed by keeping one of the components as the constant, only six independent intrinsic diffusion coefficients play a role in the development of the diffusion profiles. Similarly, when a PB couple is produced keeping two components as constant only two intrinsic diffusion coefficients play a role in the development of the diffusion profiles. Now, the question is whether an interdiffusion coefficient, let us say, $\tilde{D}^{Ni}_{CoCo}$ has the same value when estimated from different types of diffusion couples at one particular (let us say an equiatomic) composition. For this discussion, we have considered the tracer diffusion coefficients estimated by the radiotracer method [40] (see Table 5) since we need the tracer diffusion coefficient of Cr as well and since these values have a very good match with the data estimated from the PT diffusion couple method. The interdiffusion coefficient in different types of diffusion couples following Equation 13 are related by

PB Couple: $\tilde{D} = \tilde{D}^{Ni}_{CoCo} = M^B_{Co} D^{Co}_{NiNi} + M^B_{Ni} D^{Ni}_{CoCo} = M^B_{Co} D^*_{Ni} \phi^{Co}_{NiNi} + M^B_{Ni} D^*_{Co} \phi^{Ni}_{CoCo} = \frac{1}{2} D^*_{Ni} \phi^{Co}_{NiNi} + \frac{1}{2} D^*_{Co} \phi^{Ni}_{CoCo} = 2.02 \times 10^{-15}\ m^2/s$

PT Couple: $\tilde{D}^{Ni}_{CoCo} = D^{Ni}_{CoCo} - M^T_{Co}\left(D^{Ni}_{NiCo} + D^{Ni}_{CoCo} + D^{Ni}_{FeCo}\right) = D^*_{Co}\phi^{Ni}_{CoCo} - M^T_{Co}\left(\frac{M^T_{Ni}}{M^T_{Co}} D^*_{Ni}\phi^{Ni}_{NiCo} + D^*_{Co}\phi^{Ni}_{CoCo} + \frac{M^T_{Fe}}{M^T_{Co}} D^*_{Fe}\phi^{Ni}_{FeCo}\right) = D^*_{Co}\phi^{Ni}_{CoCo} - \frac{1}{3}\left(D^*_{Ni}\phi^{Ni}_{NiCo} + D^*_{Co}\phi^{Ni}_{CoCo} + D^*_{Fe}\phi^{Ni}_{FeCo}\right) = 2.11 \times 10^{-15}\ m^2/s$

Conventional couple: $\tilde{D}^{Ni}_{CoCo} = D^*_{Co}\phi^{Ni}_{CoCo} - N_{Co}\left(\frac{N_{Ni}}{N_{Co}} D^*_{Ni}\phi^{Ni}_{NiCo} + D^*_{Co}\phi^{Ni}_{CoCo} + \frac{N_{Fe}}{N_{Co}} D^*_{Fe}\phi^{Ni}_{FeCo} + \frac{N_{Cr}}{N_{Co}} D^*_{Cr}\phi^{Ni}_{CrCo}\right) = D^*_{Co}\phi^{Ni}_{CoCo} - \frac{1}{4}\left(D^*_{Ni}\phi^{Ni}_{NiCo} + D^*_{Co}\phi^{Ni}_{CoCo} + D^*_{Fe}\phi^{Ni}_{FeCo} + D^*_{Cr}\phi^{Ni}_{CrCo}\right) = 2.23 \times 10^{-15}\ m^2/s.$

It is evident from the above equations that the interdiffusion coefficient estimated from different types of diffusion couples will have a different value although the difference is small in this system. Therefore, the interdiffusion zone length will be different in different types of couples depending on the value and sign of the intrinsic diffusion coefficients. It also should be noted from Table 1 (for PT couple) that we have $\tilde{D}_{11}^{2} = \tilde{D}_{22}^{1} + \tilde{D}_{32}^{1}$ . Considering Ni, Co, Fe and Cr as the components 1, 2, 3 and 4, we have $\tilde{D}_{NiNi}^{Co} = \tilde{D}_{CoCo}^{Ni} + \tilde{D}_{FeCo}^{Ni}$. In Ni-Co PB couple, we keep Fe and Cr constant compared to Ni-Co-Fe PT couple in which Cr is kept constant. Therefore, we have $\tilde{D}_{NiNi}^{Co} = \tilde{D}_{CoCo}^{Ni}$ in a PB couple since $\tilde{D}_{FeCo}^{Ni}$ does not contribute (Fe does not produce the diffusion profile). This is indeed true (see Equation 28). However, it should be noted here again based on the discussion above that $\tilde{D}_{NiNi}^{Co}$ and $\tilde{D}_{CoCo}^{Ni}$ estimated in a PT couple will have different values compared to the values estimated in PB couples since these are related differently by the different set of intrinsic diffusion coefficients along with the difference in values of $M_i^B$ and $M_i^T$ .

Table 5 Comparison of tracer diffusion coefficients estimated following different methods. The data in PT and PB couples are calculated considering the vacancy wind effect (VWE).

| **Method** | $D^*_{Ni}$ **(×10⁻¹⁵ m²/s)** | $D^*_{Co}$ **(×10⁻¹⁵ m²/s)** | $D^*_{Fe}$ **(×10⁻¹⁵ m²/s)** | $D^*_{Cr}$ **(×10⁻¹⁵ m²/s)** |
|---|---|---|---|---|
| **PT method (considering VWE)**<br>(Ni = 24.7, Co = 24.9, Fe = 24.7, Cr = 25.7) | **2.38±0.4** | **2.37±0.4** | **7.05±1.1** | **-** |
| **PB-1 method (Ni-Co)** [4]*<br>(Ni = 24.1, Co = 24.9, Fe = 25.2, Cr = 25.8) | **2.70±0.4** | **1.90±0.3** | | |
| **PB-2 method (Fe-Cr)** [4]*<br>(Ni = 24.9, Co = 25.9, Fe = 24.4, Cr = 24.8) | | | **5.60±0.8** | **8.01±1.2** |
| **Radiotracer method** [40]<br>**Single Crystal (SC) alloy**<br>(Ni = 24.8, Co = 25.1, Fe = 25.2, Cr = 24.9) | **2.33±0.4** | **1.98±0.4** | **6.50±1.3** | **10.34±2** |
| **Radiotracer method** [41]<br>**Polycrystal (PC) alloy**<br>(Ni = 25.0, Co = 24.5, Fe = 24.7, Cr = 25.9) | **1.17±0.1** | **1.42±0.1** | **8.75±0.1** | **14.75±0.3** |

Table 6 Calculated intrinsic diffusion coefficients from the tracer diffusion coefficients neglecting and considering the vacancy wind effect (VWE) at $N_{Ni}$ = 24.7, $N_{Co}$ = 24.9, $N_{Fe}$ = 24.7, $N_{Cr}$ = 25.7 at.%. at 1200°C . The estimated intrinsic diffusion coefficients at the Kirkendall marker planes in PB diffusion couples are also listed for comparison [4] (K Plane Composition: $N_{Ni}$ = 24.1, $N_{Co}$ = 24.9, $N_{Fe}$ = 25.2, $N_{Cr}$ = 25.8 at.%).

| | **Ni dependent variable** | | | | | **Co dependent variable** | | | | | **Fe dependent variable** | |
|---|---|---|---|---|---|---|---|---|---|---|---|---|
| | **PT Method** | | **PB Method** [4] | | | **PT Method** | | **PB Method** [4] | | | **PT Method** | |
| | (neglecting VWE) | (considering VWE) | (neglecting VWE) | (considering VWE) | | (neglecting VWE) | (considering VWE) | (neglecting VWE) | (considering VWE) | | (neglecting VWE) | (considering VWE) |
| $D^{Ni}_{NiCo}$ | -2.09±0.3 | -2.10±0.3 | | | $D^{Co}_{CoNi}$ | -2.38±0.4 | -2.37±0.4 | | | $D^{Fe}_{FeNi}$ | -7.69±1.3 | -8.46±1.3 |
| $D^{Ni}_{CoCo}$ | 2.35±0.4 | 2.34±0.4 | 2.40±0.4 | 2.49±0.4 | $D^{Co}_{NiNi}$ | 2.12±0.3 | 2.12±0.3 | 1.88±0.3 | 1.82±0.3 | $D^{Fe}_{NiNi}$ | 2.52±0.4 | 2.26±0.4 |
| $D^{Ni}_{FeCo}$ | -0.35±0.06 | -0.36±0.06 | | | $D^{Co}_{FeNi}$ | 0.35±0.06 | 0.37±0.06 | | | $D^{Fe}_{CoNi}$ | 0.53±0.08 | 0.27±0.04 |
| $D^{Ni}_{NiFe}$ | -2.52±0.4 | -2.26±0.4 | | | $D^{Co}_{CoFe}$ | -2.93±0.5 | -2.66±0.5 | | | $D^{Fe}_{FeCo}$ | -8.01±1.3 | -8.87±1.3 |
| $D^{Ni}_{CoFe}$ | -0.53±0.08 | -0.27±0.04 | | | $D^{Co}_{NiFe}$ | -0.41±0.06 | -0.14±0.02 | | | $D^{Fe}_{NiCo}$ | 0.40±0.06 | 0.13±0.02 |
| $D^{Ni}_{FeFe}$ | 7.69±1.2 | 8.46±1.2 | | | $D^{Co}_{FeFe}$ | 8.11±1.3 | 8.90±1.3 | | | $D^{Fe}_{CoCo}$ | 2.89±0.4 | 2.62±0.4 |

Another important aspect should be noted here related to the estimation of the tracer diffusion coefficients from interdiffusion profiles. We cannot estimate the tracer diffusion coefficients at the cross of two PB couples in the ternary or multicomponent system since we cannot write enough independent equations. For example, in a ternary system, one PB couple might be developed by producing the diffusion profiles of components 1 and 2 keeping component 3 constant. Another PB couple might be developed by producing the diffusion profiles of components 1 and 3 keeping component 2 constant. We can write only one independent equation in a PB couple relating the interdiffusion coefficients and the tracer diffusion coefficients (see Equation 39c). Therefore, we have only two independent equations to solve for the calculation of three tracer diffusion coefficients at the cross of two PB couples in a ternary system. We, therefore, need an additional PB diffusion couple in which components 2 and 3 will also produce the ideal/near-ideal PB diffusion profile and pass through the same composition of the intersection. It further means that we should be able to produce all three possibilities of ideal/near-ideal PB diffusion couples in a ternary system passing through the same composition. We have witnessed the presence of two ideal PB couples in a composition range [3]; however, it is unlikely to find all the three combinations to produce ideal/near-ideal PB profiles in a ternary system. We are not considering the estimation of the composition-dependent diffusion coefficients from the major non-ideal PB profiles as it was done before in some other references [10, 13] for the estimation of the data without any physical significance since we cannot relate the estimated interdiffusion coefficients logically with the tracer and intrinsic diffusion coefficients by the equations applicable in ideal PB and PT diffusion couples. The clear presence of major non-ideality should be treated as a conventional diffusion couple in which all the components produce the diffusion profiles. Logically, this should be even named as the PB diffusion couple simply because the composition of a few components kept the same in the end-members. Similarly, we need four ideal or near-ideal PB couples to pass through the same composition in a four-component system if we are interested to estimate four tracer diffusion coefficients at the intersection of PB couples. It further means that there should be a possibility of producing four such diffusion couples out of 6 options (i.e. by developing diffusion profiles of components 1-2, 1-3, 1-4, 2-3, 2-4, 3-4 keeping other components as the constant). However, this is very difficult to find in most of the systems. For example, in NiCoCrFe system, only two combinations produce such diffusion profiles, as shown in different references [4, 13], which were also found in other references although in different composition ranges [14, 15]. Therefore, the intrinsic and tracer diffusion coefficients should be estimated following the Kirkendall marker experiments in a PB couple, as it is demonstrated previously [4].

Table 7 The contribution of vacancy wind effects in main and cross intrinsic diffusion coefficients in NiCoFeCr system at the cross of two PT couples.

| Ni dependent variable | | | Co dependent variable | | | Fe dependent variable | |
|---|---|---|---|---|---|---|---|
| PT Method | | PB Method [4] | PT Method | | PB Method [4] | PT Method | |
| $1+W_{NiCo}$ | 1.005 | | $1+W_{CoNi}$ | 0.996 | | $1+W_{FeNi}$ | 1.100 |
| $1+W_{CoCo}$ | 0.996 | 0.968 | $1+W_{NiNi}$ | 1.000 | 1.038 | $1+W_{NiNi}$ | 0.897 |
| $1+W_{FeCo}$ | 1.029 | | $1+W_{FeNi}$ | 1.057 | | $1+W_{CoNi}$ | 0.509 |
| $1+W_{NiFe}$ | 0.897 | | $1+W_{CoFe}$ | 0.908 | | $1+W_{FeCo}$ | 1.107 |
| $1+W_{CoFe}$ | 0.509 | | $1+W_{NiFe}$ | 0.342 | | $1+W_{NiCo}$ | 0.325 |
| $1+W_{FeFe}$ | 1.100 | | $1+W_{FeFe}$ | 1.097 | | $1+W_{CoCo}$ | 0.907 |

As already discussed earlier, the estimation of intrinsic and tracer diffusion coefficients following the Kirkendall marker experiment is almost impossible in the ternary or multicomponent system because of stringent requirement which is almost impossible to meet unless found incidentally. However, we can estimate the tracer diffusion coefficients at the cross of ternary diffusion couples in a three-component system and pseudo-ternary diffusion couples in a multicomponent system. This was for the first time proposed in a ternary system by Kirkaldy and Lane without experimental analysis because of lack of reliable thermodynamic data in that era [26]. Much later, the group of van Loo followed a similar concept for the estimation of the tracer diffusion coefficients in a ternary system [27]. Today, these analyses are relatively easier because of availability of various mathematical tools to analyze data compared to the analysis conducted by plotting the composition profiles on a chart paper during that era and the possibility of extracting thermodynamic data from Thermo-Calc or other similar sources. However, sometimes, these data are found to be not reliable in various systems when compared to the experimentally developed diffusion profiles because of the reasons described in Ref. [26].

In ternary or PT diffusion couples, we can write two independent equations from one couple and, therefore, we have four independent equations at the cross of two couples to calculate the tracer diffusion coefficients of three components which produce the diffusion profiles. If we are interested in estimating all the tracer diffusion coefficients by intersecting the PT profiles in a four-component system, we need two different sets of ideal PT profiles with a different set of components producing the diffusion profiles. This is possible to achieve, which will be reported in future considering different multicomponent systems. For example, we can estimate the tracer diffusion coefficients of components 1, 2 and 3 from intersecting PT diffusion couples in which components 1-2-3 produce the diffusion profiles keeping component 4 constant (as shown in this article). We may be able to produce another set by developing the diffusion profiles of a different combination of components including the component 4 but keeping another component (1 or 2 or 3) constant.

The same method of calculating the tracer diffusion coefficients can be extended to the body diagonal diffusion couples also if one is ready to compromise the strict restriction of estimating the diffusion coefficients at the intersection of diffusion couples. When the diffusion couples are produced by developing the diffusion profiles of all the components, we need three diffusion couples in a four-component system and four diffusion couples in a five-component system to intersect at one composition i.e. at one point in multicomponent space which is very difficult to achieve even in small composition range since the diffusion coefficients are not expected to be constant exactly [11]. With the increasing number of components, the required number of diffusion couples to intersect increase gradually making this method even difficult. There is a possibility that only two or three couples pass through closely in a set of the required number of diffusion couples in a multicomponent system. In such a scenario, an estimation of the tracer diffusion coefficients initially from only two diffusion couples directly from the interdiffusion flux instead of estimating the interdiffusion coefficients will make the effort easier only if relatively reliable thermodynamic data can be generated from Thermo-Calc or similar sources. Following, one can estimate the intrinsic and interdiffusion coefficients from the estimated tracer diffusion coefficients.

To explain this further, let us consider an n-component system. In a conventional method when all the components develop the diffusion profiles, the intrinsic flux of component *i* neglecting the vacancy wind effect can be expressed as (see supplementary file)

$$J_i = -L_{ii}\frac{\partial \mu_i}{\partial x} = \frac{C_i D_i^*}{RT}\frac{\partial \mu_i}{\partial x} = -\frac{N_i D_i^*}{V_m RT}\frac{\partial \mu_i}{\partial x} \tag{44a}$$

Replacing Equation 10c, we can write

$$J_i = -\frac{N_i D_i^*}{V_m RT}\left(\frac{\partial \mu_i}{\partial N_2}\frac{\partial N_2}{\partial x} + \frac{\partial \mu_i}{\partial N_3}\frac{\partial N_3}{\partial x} + \cdots \frac{\partial \mu_i}{\partial N_n}\frac{\partial N_n}{\partial x}\right) = -\frac{N_i D_i^*}{V_m RT}\left(\frac{RT}{N_2}\emptyset_{i2}^1\frac{\partial N_2}{\partial x} + \frac{RT}{N_3}\emptyset_{i3}^1\frac{\partial N_3}{\partial x} \ldots + \frac{RT}{N_n}\emptyset_{in}^1\frac{\partial N_n}{\partial x}\right) \tag{44b}$$

Following, these are related to the interdiffusion flux by

$$\tilde{J}_i = J_i - N_i\sum_{i=1}^{n} J_i \tag{45a}$$

where we have $\sum_{i=1}^{n}\tilde{J}_i = 0$ i.e. $\tilde{J}_1 = -\sum_{i=2}^{n}\tilde{J}_i$ (45b)

Replacing Equation 44b in Equation 45a, we can express (*n-1*) independent equations in a *n* component system for the interdiffusion fluxes by relating *n* tracer diffusion coefficients in a diffusion couple. Therefore, we can estimate these *n* number of tracer diffusion coefficients from only two diffusion profiles by writing *2(n-1)* independent equations. Irrespective of the number of components (in a system with three or more components), we need only two diffusion couples either to intersect or, if ready to compromise, when the profiles pass close to each other (let us say within 0.5-1 at.%, which should be established based on error analysis). Following, we can estimate the intrinsic and interdiffusion coefficients utilizing the calculated tracer diffusion coefficients using the relations expressed in the supplementary file. One can follow the same even by considering the vacancy wind effect. This may make the effort relatively easier since otherwise the need for (*n*-1) diffusion profiles in a multicomponent system to intersect or pass through closely make the task difficult considering the fact that these should be established based on trial experiments.

**4. Conclusion**

We have demonstrated the estimation of the tracer and intrinsic diffusion coefficients from PT diffusion couples. This is demonstrated in NiCoFeCr system by comparing with the tracer diffusion coefficients estimated by the radiotracer method. The outcome of this study can be summarized as:

- The estimation of the interdiffusion coefficients by intersecting the PT diffusion couples were demonstrated earlier [3, 4]. In this study, we have proposed the relations for calculating the tracer and intrinsic diffusion coefficients at the composition of the intersection. These are written for ideal/near-ideal diffusion profiles i.e. when the profiles of the components which supposed to remain constant do not show uphill nature. This can also be used in non-ideal diffusion profiles if the non-ideality is minor in nature, which can be considered by comparing the composition gradients of the components. Error will be negligible if the composition gradients of the components which are desired to remain constant are very small compared to composition gradients of the components which are desired to develop the diffusion profiles. A detailed analysis on these aspects will be published later based on experimental results with different ranges of non-ideality in various systems.
- In the process of developing the relations in PT diffusion couples, we have shown a link between the relations proposed by Onsager, Kirkaldy and Manning in different types of diffusion couples. This helps to understand the underlying diffusion process in different types of situations in a multicomponent system.
- The calculated tracer diffusion coefficients from the PT diffusion couples are found to have a good match with the tracer diffusion coefficients estimated from the PB diffusion couples at the Kirkendall marker plane and estimated following the radiotracer method.
- Following, the calculated intrinsic diffusion coefficients help to understand the influence of these parameters on the interdiffusion coefficients in relation to the thermodynamic parameters. Therefore, one can now understand the diffusional interactions between the components based on the calculated intrinsic diffusion coefficients, which was not possible earlier. We have further demonstrated that the intrinsic diffusion coefficients are the same (if the vacancy wind effect is negligible) but the interdiffusion coefficients are different when estimated from different types of diffusion couples. However, a few cross intrinsic diffusion coefficients are found to be strongly affected by the vacancy wind effect in PT couples. The same is expected to be found when all the components produce the diffusion profiles.
- We also have proposed the estimation of the tracer diffusion coefficients from only two diffusion profiles in conventional diffusion couple method in the multicomponent system. This will facilitate the estimation of all types of diffusion coefficients if two diffusion couples could be produced passing very closely (if not intersected) following the concept of the body diagonal method. This is easier than the requirement of (n-1) diffusion couples to intersect or pass closely for the direct estimation of the interdiffusion coefficients. This can be a method of choice in a system with a higher number of components if reliable thermodynamic parameters are available.
- We should target for estimating both intrinsic and tracer diffusion coefficients considering the vacancy wind effect in a multicomponent system from PT couples. Numerical approach implemented in software packages like DICTRA do not consider the vacancy wind effect although it may play an important role in multicomponent diffusion.

- Maintaining the facility for radiotracer method is not easy and, therefore, only very few groups are currently following this method for estimation of the tracer diffusion coefficients. Experiments with radioisotopes of various elements are not feasible because of various reasons such as high cost and short half-life (for example, Al, Ga and Si). On the other hand, the diffusion couple method is easier to follow and therefore common. As demonstrated in this article, the newly proposed PB and PT methods can be very useful first to estimate the tracer diffusion coefficients of the components and then estimate all types of diffusion coefficients, which are otherwise difficult to estimate following the conventional diffusion couple method i.e. by producing diffusion profile of all the components.

**Acknowledgements:** We would like to acknowledge the financial support from ARDB, India, Grant number: ARDB/GTMAP/01/2031786/M. AP acknowledges the discussion with Prof. Kaustubh Kulkarni, IIT Kanpur, India.